\documentclass[useAMS,usenatbib]{mn2e}
\usepackage{graphicx}
\usepackage{color}
\usepackage{soul}
\usepackage{bm}
\usepackage{amssymb,amsfonts,amsmath}%

\newcommand{\bec}[1]{\mbox{\boldmath $ #1$}}

\newcommand{\meanrho}{\overline{\rho}}

{}
{}
{}

\newcommand{\meanAAA}{\overline{\mathsf{A}}}

{}
{}
{}
{}
{}
{}
{}
{}
\newcommand{\meanBB}{\overline{\mbox{\boldmath $B$}}{}}{}
{}
{}
{}
{}
{}
{}
{}
{}

{}
{}
{}
\newcommand{\meanA}{\overline{A}}
\newcommand{\meanB}{\overline{B}}

\title[Magnetic fields of low-mass main sequences stars]
{Magnetic fields of low-mass main sequences stars: Non-linear dynamo theory
and mean-field numerical simulations}

\author[N. Kleeorin,  I. Rogachevskii, N. Safiullin, R. Gershberg, S. Porshnev]
{ N. Kleeorin,$^{1,2}$
  I. Rogachevskii,$^{1,3}$
  N. Safiullin,$^{4,2}$
  R. Gershberg,$^{5}$
  S. Porshnev$^{4,6}$ \\
 $^{1}$ Department of Mechanical Engineering,
        Ben-Gurion University of Negev, POB 653, 84105 Beer-Sheva, Israel\\
 $^{2}$ Institute of Continuous Media Mechanics, Korolyov str. 1, Perm  614013, Russia\\
 $^{3}$ Nordita, KTH Royal Institute of Technology and Stockholm University,
        Roslagstullsbacken 23, SE-10691 Stockholm, Sweden\\
 $^{4}$ Department of Radio Electronic and Informational Technology,
        Ural Federal University, 19 Mira str., 620002 Ekaterinburg, Russia\\
 $^{5}$ Crimean Astrophysical Observatory, RAN, 298409 Nauchny, Russia\\
 $^{6}$ N.N. Krasovskii Institute of Mathematics and Mechanics (IMM UB RAS), 620108 Ekaterinburg, Russia}

\begin{document}


\maketitle


\begin{abstract}
Our theoretical and numerical analysis have suggested that for low-mass main sequences stars (of the spectral classes from M5 to G0)
rotating much faster than the Sun, the generated large-scale magnetic field is caused by the mean-field $\alpha^2\Omega$ dynamo,
whereby the $\alpha^2$ dynamo is modified by a weak differential rotation.
Even for  a weak differential rotation, the behaviour of the magnetic activity is changed drastically
from aperiodic regime to non-linear oscillations and appearance of a chaotic behaviour
with increase of the differential rotation.
Periods of the magnetic cycles decrease with increase of the differential rotation,
and they vary from tens to thousand years.
This long-term behaviour of the magnetic cycles may be related to the characteristic time
of the evolution of the magnetic helicity density of the small-scale field.
The performed analysis is based on the mean-field simulations (MFS) of the $\alpha^2\Omega$ and $\alpha^2$ dynamos
and a developed non-linear theory of $\alpha^2$ dynamo.
The applied MFS model was calibrated using turbulent parameters
typical for the solar convective zone.
\end{abstract}

\maketitle


\section{Introduction}

The cold dwarf stars of the main sequences of the spectral class M composing 70 - 75\% of
all star population, have smaller sizes ($0.1 R_\odot < R < 0.8 R_\odot$)
in comparison with the Sun, smaller masses $(0.08 M_\odot < M < 0.55 M_\odot)$,
smaller luminosity $(L \leq 0.05 L_\odot)$ and effective temperatures of 2500–4000 K,
where $R_\odot$, $M_\odot$ and $L_\odot$ are the solar radius, mass and luminosity,
respectively \citep[see, e.g.,][]{BHC10,PM13,WHJ19,KO21}.
About 15 -- 20 \% of these stars have similar magnetic activity as the Sun with cold magnetic spots
and sporadic flares of very high releasing energy in the form of radiations
in wide range of wavelength including thermal and nonthermal X-ray \citep{HDK14,NIC17}.
As the Sun, these stars obey differential rotation and have similar atmospheric structure, consisting
of photosphere, hot chromosphere and corona \citep{WNW18,GKP20}.

According to various observations \citep[see, e.g.,][]{SL85,SA96,DMP03,DMP08,RB07},
slow rotating stars $(\Omega < \Omega_\odot)$
have values and structures of the large-scale magnetic field similar to
solar magnetic field, where $\Omega_\odot$ is the solar angular velocity.
On the other hand, fast rotating stars $(\Omega > 10\, \Omega_\odot)$
have strong poloidal magnetic fields at the pole, and sometimes they have
strong toroidal magnetic fields at the pole \citep{ST09,MDP10}.
The periods of the stellar cycles can be in several times larger than the periods of the solar cycles \citep{BKL19}.
Magnetic fields of fast rotating stars can be more than several thousands Gauss \citep{KHL20,KO21}.

Various MHD direct numerical simulations (DNS) and large eddy simulations (LES)
of convection and dynamos in low-mass convective stars have been performed
in a number of studies \citep[see, e.g.,][]{DSB06,BROW08,YCW16,BOV20,KA21,BT22}.
They use fully compressible MHD system with weak density stratification or anelastic simulations
with strong density stratification in a box or spherical shell.
Main results of these simulations are summarised in review by \cite{KBB23}.
In particular, when the magnetic field is weak or absent,
both “solar” and “anti-solar” differential rotation can be formed.
When the dynamo generated large-scale magnetic field is strong, it
reduces the differential rotation
sometimes resulting to the solid-body rotation.
The dynamo generated large-scale magnetic field
is mainly axisymmetric, and it has the
dipole or quadrupole structure depending on rotation, shear, and density
stratification.
In particular, when rotation is strong and shear is weak, the magnetic field has dipolar structure
\citep{GDW12,SPD12,YCM15a}.
There are many simulations with highly-stratified, vigorous convection
that also show dipole magnetic structure \citep{YCM15a}.
In the presence of large-scale shear, propagating dynamo waves
are observed in simulations \citep{YCW16,KA21,BT22}.
Some simulations also produce non-axisymmetric magnetic field
\citep{KA21,BT22}.

Various mean-field dynamo models have been suggested to explain
generation of large-scale magnetic fields in M dwarfs
\citep[see, e.g.,][]{CK06,KMS14,SSK15,PI17,PY18}.
In particular, mean-field simulations (MFS) of the $\alpha^2$ dynamo have been performed by \cite{CK06}.
They consider a fully convective rotating star and focus on the kinematic dynamo problem.
The large-scale magnetic field is excited when the Coriolis number ${\rm Co} = 2 \Omega_\ast \tau \geq 1$, and
the dynamo generates a non-axisymmetric steady magnetic field that is symmetric with respect to the equatorial plane \citep{CK06}.

\cite{KMS14} suggest that M-dwarfs have two types of magnetic activity: (i) magnetic cycles with
strong (kilogauss) almost axisymmetric poloidal magnetic fields;
and (ii) considerably weaker non-axisymmetric fields with a substantial toroidal component
observed at times of magnetic field inversion.
To show this, they use a kinematic model of an $\alpha^2\Omega$ dynamo with
the differential rotation determined using the numerical mean-field model by \cite{KO11}.
Applying this model, they study a magnetic field evolution
and find a transition from steady to oscillatory dynamos with increasing
turbulent magnetic Prandtl number.
Using this approach, \cite{SSK15} suggest four magnetic configurations that appear
relevant to dwarfs from the viewpoint of the dynamo theory, and
discuss observational tests to identify the configurations observationally.

\cite{PI17} has performed mean-field numerical simulations with the non-linear axisymmetric
and non-axisymmetric $\alpha^2\Omega$ dynamos
of the fully convective star with the mass $M=0.3M_\odot$
rotating with a period of 10 days.
The differential rotation is determined using the numerical mean-field model similar to \cite{KO11}.
This dynamo model also includes the meridional circulation, while
the magnetic feedback on the non-axisymmetric flows is neglected.
The dynamical quenching of the $\alpha$ effect
is described by equation for the total magnetic helicity density.
These mean-field numerical simulations yield different dynamo solutions
depending on parameters, including variations of the turbulent magnetic Prandtl number,
as a key parameter. Increase of this parameter increases
the period of the magnetic cycles.

The effects of the cross-helicity in the full-sphere large-scale mean-field dynamo models
have been studied by \cite{PY18} in the absence of the differential rotation.
They found that non-axisymmetric magnetic field
is generated when the cross-helicity and the $\alpha$ effect operate independently of each other,
while their joint action generates preferably axisymmetric dipole magnetic fields.

In the present theoretical study and mean-field numerical simulations,
we show that for the main sequences low-mass fast rotating stars,
the generated large-scale magnetic field is due to
the mean-field $\alpha^2\Omega$ dynamo, in which
the $\alpha^2$ dynamo is modified by a weak differential rotation.
This implies that for this mean-field dynamo
$R_\omega \ll R_{\alpha} R_{\alpha}^{\rm cr}$,
where $R_\omega = (\delta \Omega) \, R_\ast^2 / \eta_{_{T}}$ and
$R_{\alpha} = \alpha_\ast R_\ast / \eta_{_{T}}$ are the key dimensionless parameters
characterising the mean-field $\alpha^2\Omega$ dynamo instability, and $R_{\alpha}^{\rm cr}$
is the threshold required for the excitation of  the mean-field dynamo instability,
defined by the conditions $\gamma=0$ and $R_\omega=0$.
Here $R_\ast$ is the star radius,
$\eta_{_{T}}$ is the turbulent magnetic diffusion coefficient,
$\delta\Omega$ is the differential rotation,
$\alpha_\ast$ is  the maximum value of the
kinetic $ \alpha $ effect, and $\gamma$ is the dynamo growth rate.

We find that periods of the magnetic activity cycles decrease with increase of the differential rotation,
and they can vary from tens to thousand years.
The dynamical quenching of the $\alpha$ effect
due to the evolution of the magnetic helicity density of the small-scale field,
determines a long-term behaviour of the magnetic cycles.

This paper is organized as follows.
In Sec.~\ref{sect2} we consider the radial profiles of turbulent parameters
in a the stellar convective zones, and discuss
the theoretical rotating profiles of the kinetic $\alpha$ effect obtained using theory of
the convecting rotating MHD turbulence \citep{KR03}.
In Sec. ~\ref{sect3} we study mean-field $\alpha^2 \, \Omega$ dynamo, where
we start with the kinematic $\alpha^2 \, \Omega$ dynamo (Sec. ~\ref{subsect3.1}),
discuss the algebraic and dynamic non-linearities  (Sec. ~\ref{subsect3.2}),
and continue with
mean-field numerical simulations of the $\alpha^2\Omega$ dynamo (Sec. ~\ref{subsect3.3})  as
well as the $\alpha^2$  dynamo (Sec. ~\ref{subsect3.4}).
In Sec. ~\ref{sect4} we develop non-linear theory of axisymmetric $\alpha^2$ dynamo.
Finally,  in Sec.~\ref{sect5} we discuss the obtained results,
compare various numerical models and outline conclusions.

\section{Radial profiles of turbulent parameters in a the stellar convective zones}
\label{sect2}

In this Section, we discuss radial profiles of various turbulent parameters
in a the stellar convective zones.
As a turbulent model of stellar convective zones, we use ''Modules for Experiments
in Stellar Astrophysics (MESA)'' \citep{PB11}.
The MESA (http://mesa.sourceforge.net/)
is one-dimensional stellar evolution module, which
combines many of the numerical and physics modules
for simulations of a wide range of stellar evolution scenarios
ranging from very low mass to massive stars.

The MESA includes a module which implements the standard mixing length
theory (MLT) of convection \citep{CG68}, as well as the modified MLT \citep{HVB65}.
Whereas the standard MLT assumes high optical
depths and no radiative losses, the modified MLT
allows the convective efficiency to vary with the opaqueness of
the convective element, which is an important effect for convective zones
near the outer layers of stars \citep{HVB65}.

\begin{figure}
\centering
\includegraphics[width=8.5cm]{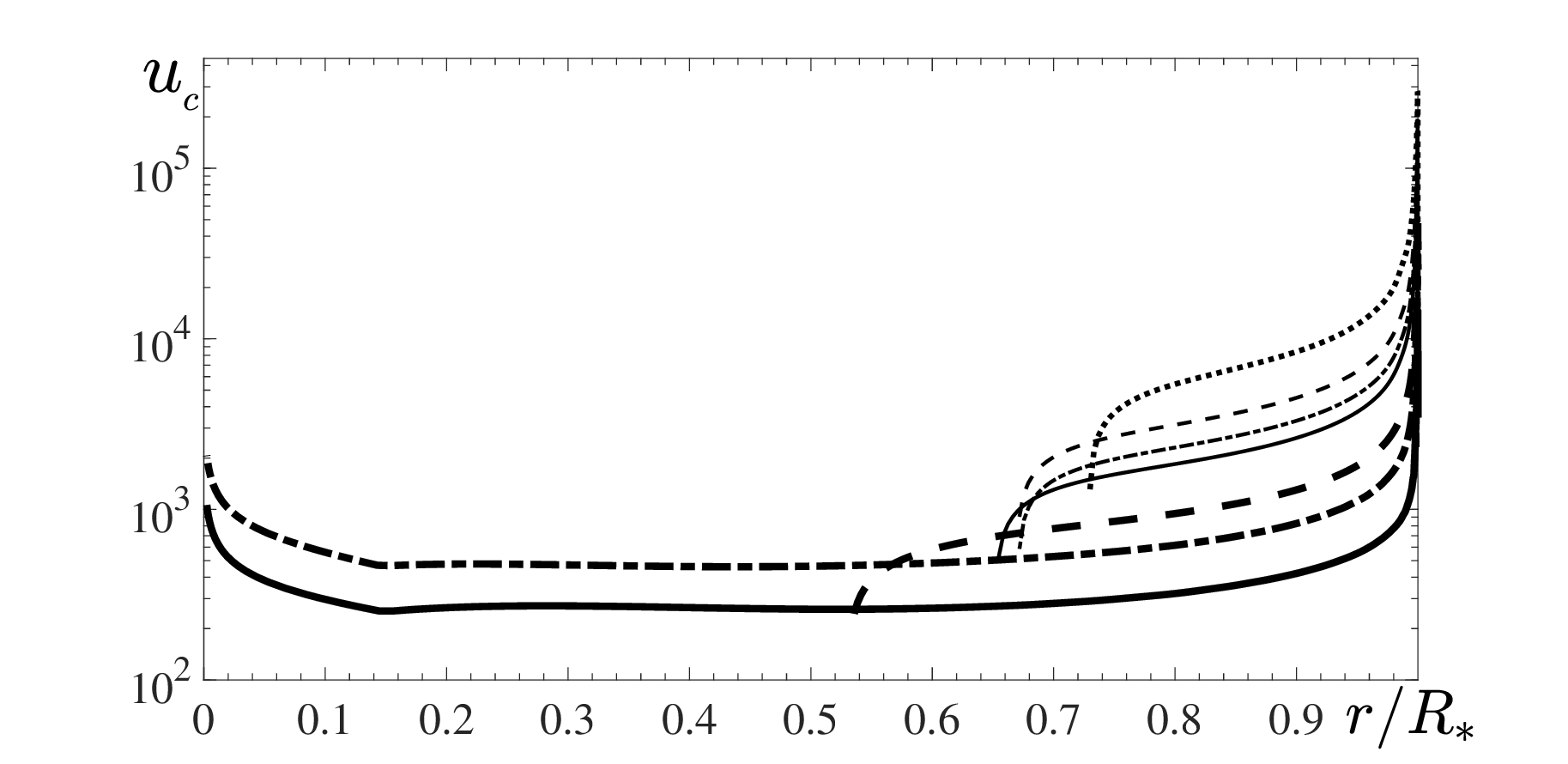}
\caption{\label{Fig1}
The radial profiles of the convective turbulent velocity $u_c$
for the main sequences stars of the spectral classes: M6 (thick solid); M4 (thick dashed-dotted);
M2 (thick dashed); K7  (thin solid); K4 (thin dashed-dotted); K2 (thin dashed);
G2 (thin dotted). The velocity is measured in cm/s.
Here $R_\ast$ is the star radius.
}
\end{figure}

\begin{figure}
\centering
\includegraphics[width=8.5cm]{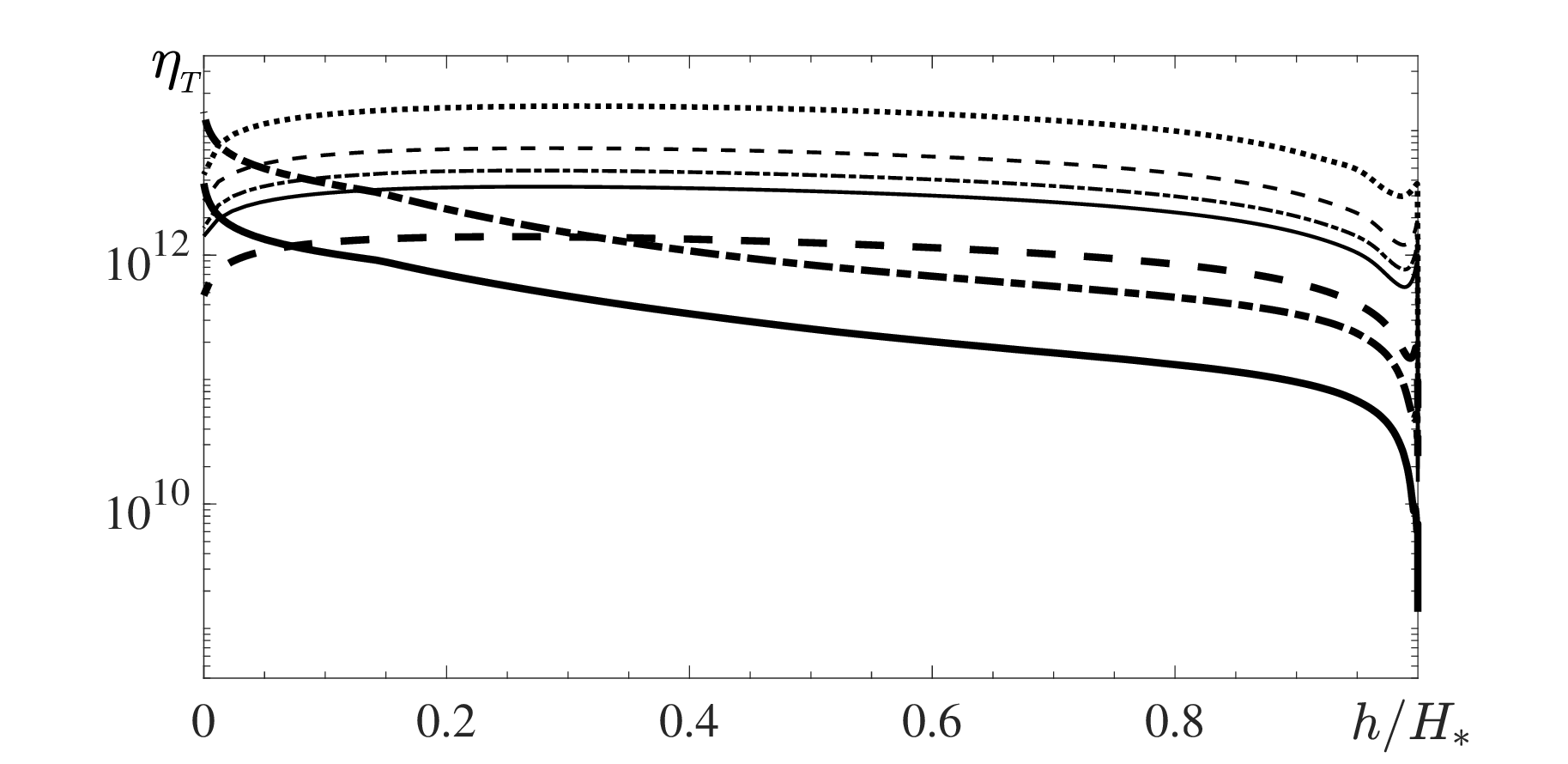}
\caption{\label{Fig2}
The radial profiles of the turbulent magnetic diffusivity $\eta_{_{T}}$
for the main sequences stars of the spectral classes: M6 (thick solid); M4 (thick dashed-dotted);
M2 (thick dashed); K7  (thin solid); K4 (thin dashed-dotted); K2 (thin dashed);
G2 (thin dotted). The turbulent magnetic diffusivity is measured in cm$^2$/s.
Here $H_\ast$ is the thickness of the convective zone, and $h$ is the hight from
the bottom of the convective zone.
}
\end{figure}

\begin{figure}
\centering
\includegraphics[width=8.5cm]{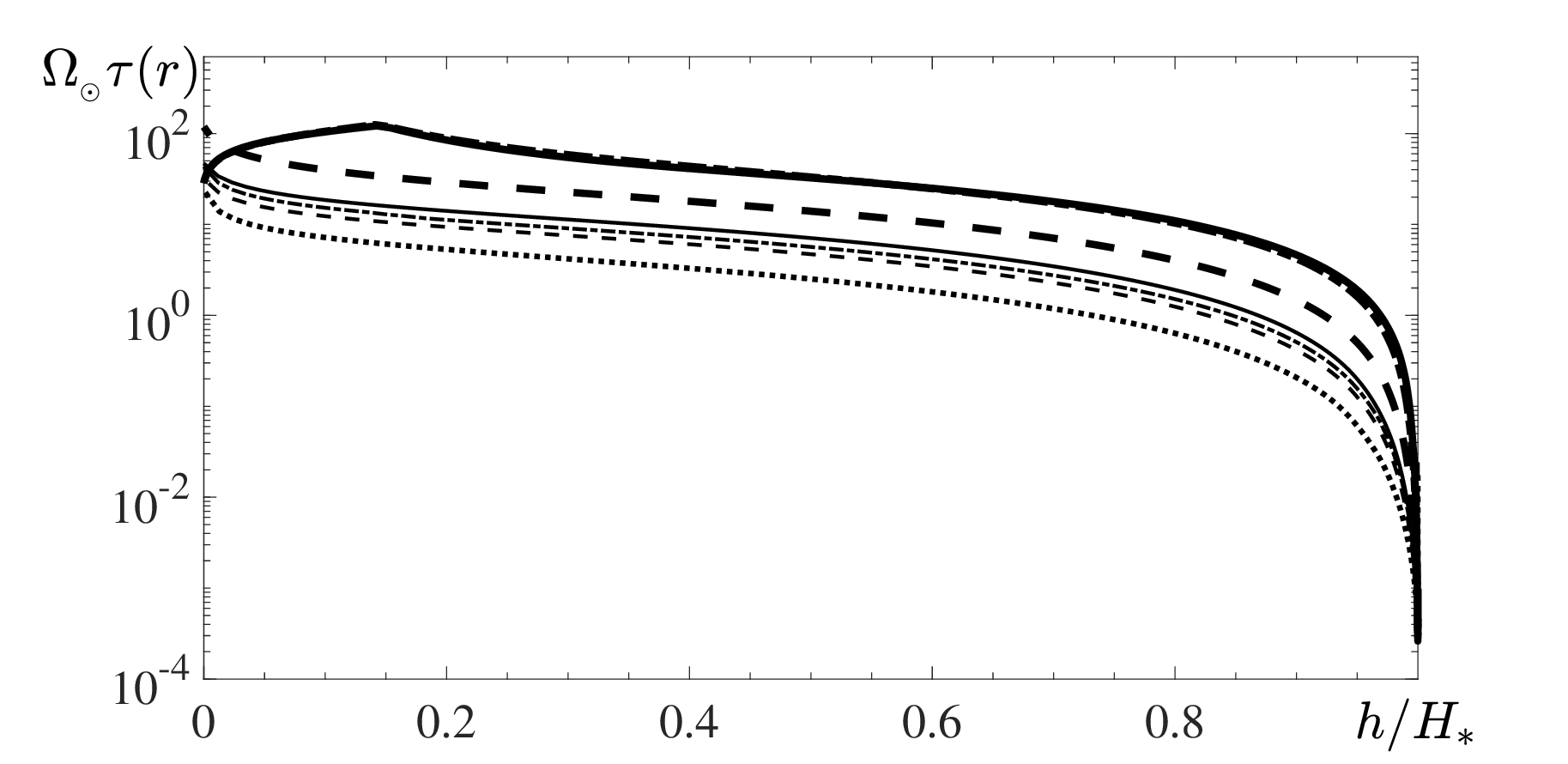}
\caption{\label{Fig3}
The radial profiles of the Coriolis parameter $\Omega_\odot \tau(h)$
based on the solar angular velocity $\Omega_\odot$ and
the turbulent turn-over time $\tau(r)=3\eta_{_{T}}/u_c^2$
for the main sequences stars of the spectral classes: M6 (thick solid); M4 (thick dashed-dotted);
M2 (thick dashed); K7  (thin solid); K4 (thin dashed-dotted); K2 (thin dashed);
G2 (thin dotted). The turbulent magnetic diffusivity is measured in cm$^2$/s.
}
\end{figure}

Using the MESA, we plot the radial profiles of the convective turbulent velocity $u_c$ (Fig.~\ref{Fig1}),
the turbulent magnetic diffusivity $\eta_{_{T}}$ (Fig.~\ref{Fig2})
and the Coriolis parameter $\Omega_\odot \tau(r)$ (Fig.~\ref{Fig3})
based on the solar angular velocity $\Omega_\odot$ and
the turbulent turn-over time $\tau(r)=3\eta_{_{T}}/u_c^2$
for stars of late spectral classes: M6, M4, M2, K7, K4, K2,
and G2. Here $H_\ast$ is the thickness of the convective zone, $h$ is the hight from
the bottom of the convective zone,
the velocity is measured in cm/s and $R_\ast$ is the star radius.
Depending on the spectral class and the depth of the convective zone,
the convective turbulent velocity $u_c$
changes from $10^2$ cm/s to $10^5$ cm/s.
Strong changes in $u_c$ occur in the upper part of the convective zone  (see Fig.~\ref{Fig1}).

On the other hand, the turbulent magnetic diffusivity $\eta_{_{T}}$ varies inside
the convective zone only in several times for stars of the spectral classes from M2 to G2,
while it changes by two orders of magnitude for stars of the spectral classes from M4 and M6 (see Fig.~\ref{Fig2}).
The Coriolis parameter $\Omega_\odot \tau(r)$
based on the solar angular velocity $\Omega_\odot$ and
the turbulent turn-over time $\tau(r)=3\eta_{_{T}}/u_c^2$
strongly decreases from $10^{2}$ near the base of the convective zone to $10^{-2} - 10^{-4}$ near the star surface
depending on the spectral class (see Fig.~\ref{Fig3}).
Note that models of the solar convective zone are given by \cite{BT66,S74}.

\begin{figure}
\centering
\includegraphics[width=8.5cm]{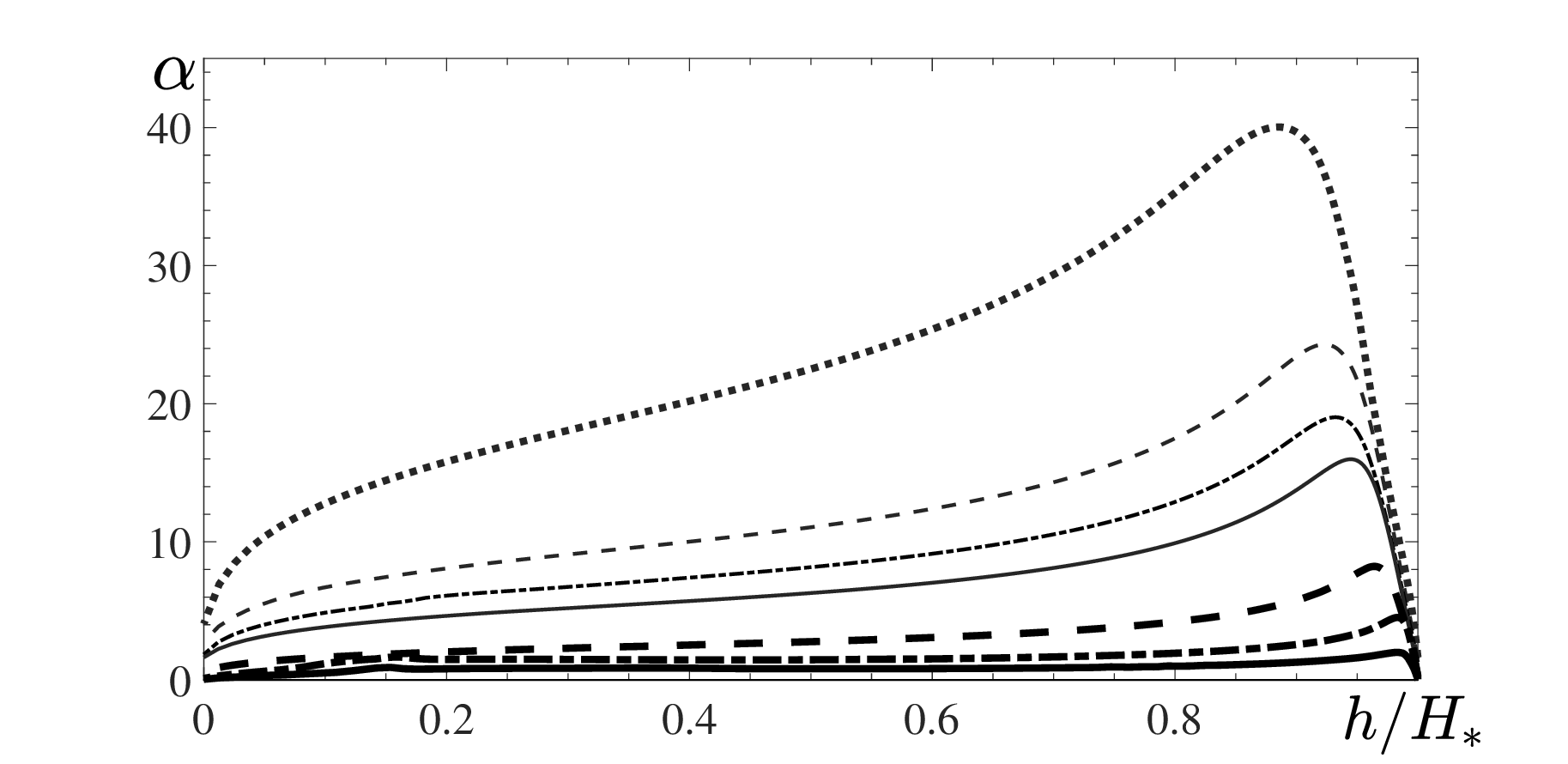}
\caption{\label{Fig4}
The radial profiles of the kinetic $\alpha$  effect
at the pole (at the  latitude $\phi=\pi/2$) for isotropic turbulent convection for $\sigma=1$ and $\varepsilon=0$
for the main sequences stars of the spectral classes: M6 (thick solid); M4 (thick dashed-dotted);
M2 (thick dashed); K7  (thin solid); K4 (thin dashed-dotted); K2 (thin dashed);
G2 (thick dotted). The kinetic $\alpha$ is measured in m/s.
}
\end{figure}

\begin{figure}
\centering
\includegraphics[width=8.5cm]{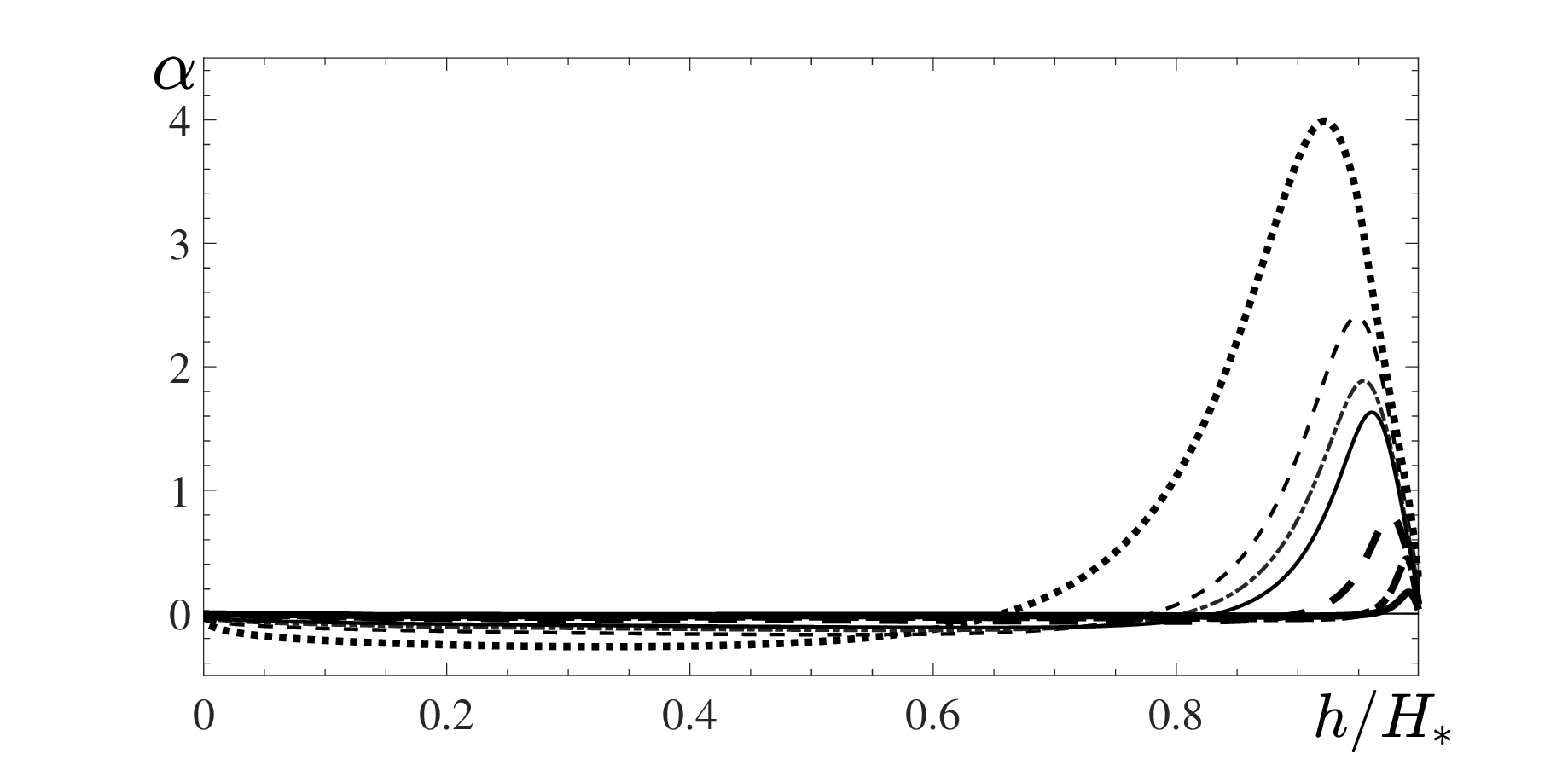}
\caption{\label{Fig5}
The radial profiles of the kinetic $\alpha$  effect
at the  latitude  $\phi=\pi/6$ for anisotropic turbulent convection for $\sigma=2$ and $\varepsilon=1.2$
for the main sequences stars of the spectral classes: M6 (thick solid); M4 (thick dashed-dotted);
M2 (thick dashed); K7  (thin solid); K4 (thin dashed-dotted); K2 (thin dashed);
G2 (thick dotted). The kinetic $\alpha$ is measured in m/s.
}
\end{figure}

\begin{figure}
\centering
\includegraphics[width=8.5cm]{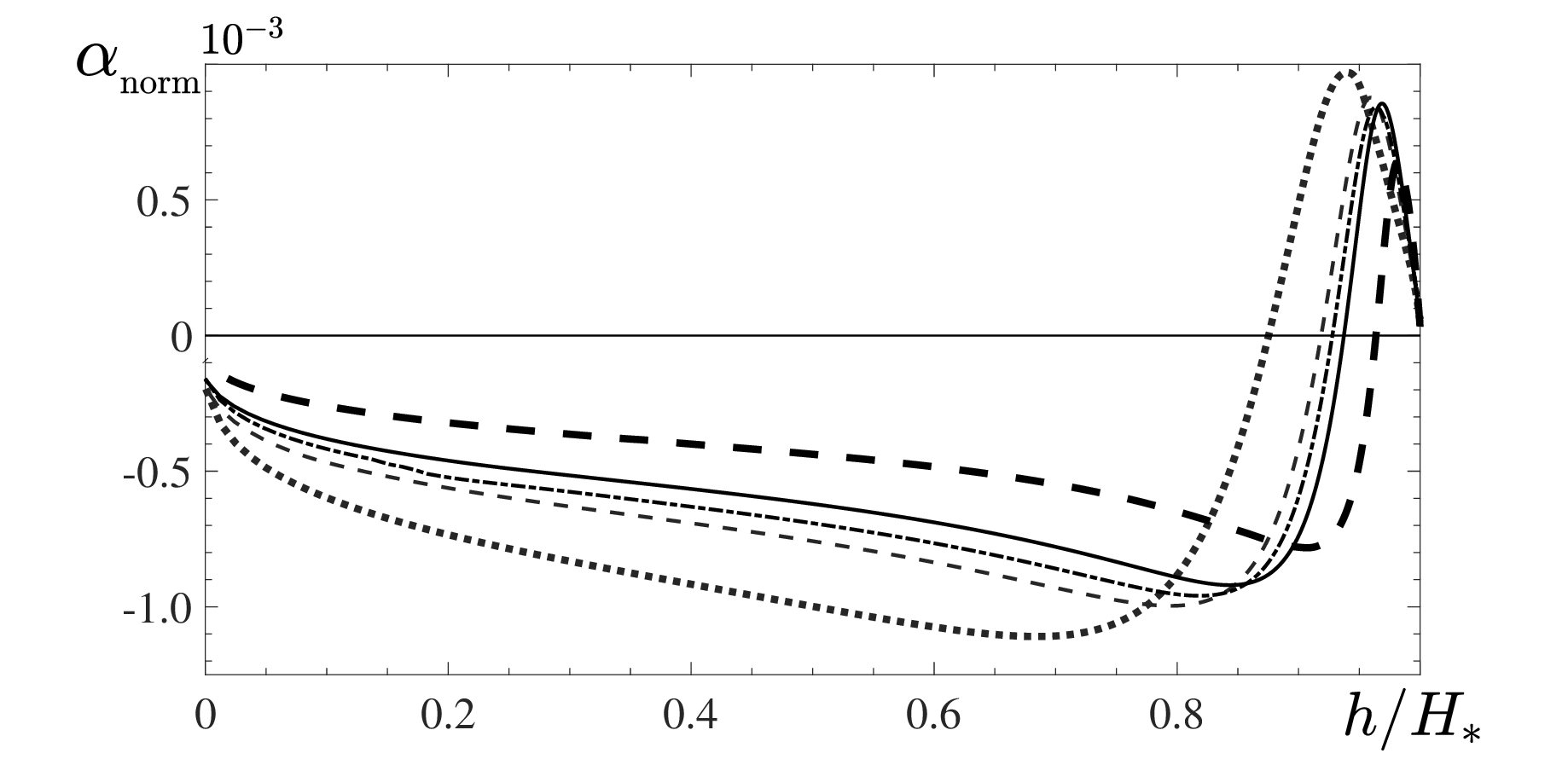}
\caption{\label{Fig6}
The radial profiles of the normalized kinetic $\alpha$ effect,
$\alpha_{\rm norm} =\alpha/(\Omega_\odot\, \eta_{_{T}})^{1/2}$,
at the pole ($\phi=\pi/2$) for $\sigma=2$ and $\varepsilon=1.2$,
for the main sequences stars with the solar rotation rate of the spectral classes:
M6 (thick solid); M4 (thick dashed-dotted); M2 (thick dashed);
K7  (thin solid); K4 (thin dashed-dotted); K2 (thin dashed);
G2 (thick dotted).
Here $H_\ast$ is the hight of the convective zone, and $h$ is the hight from
the bottom of the convective zone.
}
\end{figure}

\begin{figure}
\centering
\includegraphics[width=8.5cm]{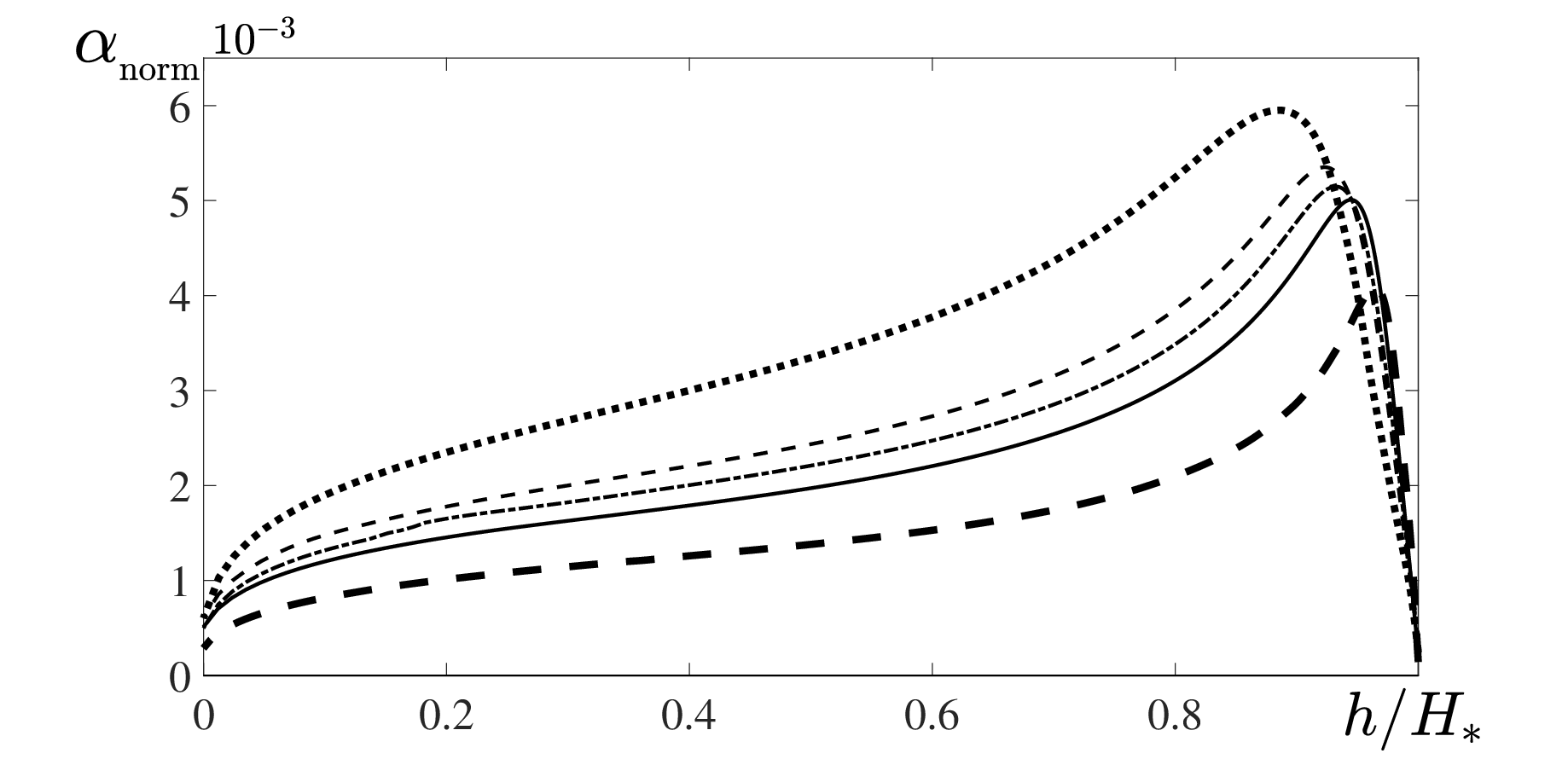}
\caption{\label{Fig7}
The radial profiles of the normalized kinetic $\alpha$ effect,
$\alpha_{\rm norm} = \alpha/(10\Omega_\odot\, \eta_{_{T}})^{1/2}$,
at the pole ($\phi=\pi/2$) for $\sigma=1$ and $\varepsilon=0$,
for the main sequences stars  of the spectral classes: M6 (thick solid); M4 (thick dashed-dotted);
M2 (thick dashed); K7  (thin solid); K4 (thin dashed-dotted); K2 (thin dashed);
G2 (thick dotted).
Here $H_\ast$ is the hight of the convective zone, and $h$ is the hight from
the bottom of the convective zone.
}
\end{figure}

Models of the stellar convective zones based on the standard mixing length
theory do not take into account the effect of the Coriolis force
on the convective turbulence.
One of the key effects of rotation in density-stratified convection is
\begin{itemize}
\item{
production of the kinetic helicity and the kinetic $\alpha$ effect,}
\item{
formation of the differential rotation and}
\item{
strong anisotropization of turbulence.}
\end{itemize}
Using the results obtained applying the theory of the convecting
rotating MHD turbulence \citep{KR03,BGKR13}, we plot in Figs.~\ref{Fig4}--\ref{Fig7}
the isotropic part of the kinetic ${\bm \alpha}$ tensor
that characterises the kinetic ${\bm \alpha}$ effect,
\begin{eqnarray}
\alpha &=& {1 \over 6} \biggl({\ell_{0}^{2} \Omega \over
H_{\rho}} \biggr) \sin \phi \, \left[\Psi_{1}(\omega) +
\Psi_{2}(\omega) \sin^{2} \phi\right] ,
\label{F1}
\end{eqnarray}
where $\phi$ is the latitude, $\Omega$ is the angular velocity,
$H_{\rho}$ is the density stratification hight,
$\ell_{0}$ is the integral scale of turbulent convection,
the parameter $\omega=4 \Omega \tau(r)$,
and the functions $\Psi_{1}(\omega)$ and $\Psi_{2}(\omega)$
are given in Appendix~\ref{appendix-A}.

For instance, for a slow rotation $(\omega \ll 1)$,
the kinetic $\alpha$ effect is given by
\begin{eqnarray}
\alpha = {4 \over 5} \biggl({\ell_{0}^{2} \Omega \over
H_{\rho}} \biggr) \biggl(2 - {\sigma \over 3} - {5 \lambda \over
6} \biggr) \sin \phi \;,
\label{F14}
\end{eqnarray}
and for fast rotation $(\omega \gg 1)$ it is given by
\begin{eqnarray}
\alpha &=& - {\pi \over 32} \biggl({\ell_{0} u_{c} \over
H_{\rho}} \biggr) \biggl(2 \lambda + {\sigma \over 3} - 3
+ (\sigma - 1) \sin^2 \phi \biggr) \sin \phi ,
\nonumber\\
\label{F15}
\end{eqnarray}
where $u_{c}=\ell_{0} / \tau$ is the characteristic turbulent velocity.
Here the parameter $\lambda=2 \varepsilon/(\varepsilon+2)$ is
related to the degree of anisotropy $\varepsilon$
of turbulent velocity field:
\begin{eqnarray}
\varepsilon = {2 \over 3} \biggl( {\langle {\bf u}_{\perp}^{2}
\rangle \over \langle {\bf u}_{z}^{2} \rangle} - 2\biggr),
\label{FF1}
\end{eqnarray}
${\bf u}_{\perp}$ is the horizontal turbulent velocity,
${\bf u}_{z}$ is the vertical turbulent velocity
(in the direction of gravity).
The parameter $\sigma$ determines the degree of thermal anisotropy.
For $\sigma < 1$, the thermal plumes in a convective turbulence have the form
of column or thermal jets, while for $\sigma > 1$, the
`'pancake'' thermal plumes exist in the background turbulent convection.

For example, in  Fig.~\ref{Fig4} we show
the radial profiles of the kinetic $\alpha$  effect
at the pole (at the  latitude $\phi=\pi/2$) for isotropic turbulent convection for $\sigma=1$ and $\varepsilon=0$,
while in Fig.~\ref{Fig5} we plot the kinetic $\alpha$  effect
at the  latitude  $\phi=\pi/6$ for anisotropic turbulent convection for $\sigma=2$ and $\varepsilon=1.2$.
Various curves in Figs.~\ref{Fig4} and~\ref{Fig5}
correspond to stars of the spectral classes from M6 to G2.
We use here the radial profile of the Coriolis parameter $\Omega_\odot \tau(r)$
to determine the radial profile of the kinetic ${\bm \alpha}$ effect
which is the function of the Coriolis parameter.
It follows from Figs.~\ref{Fig4} and~\ref{Fig5} that the maximum value $\alpha_\ast$
of the kinetic $\alpha$  effect depends on the spectral class at a given rotation rate.
For instance, the stars of the spectral class G2 have largest values of $\alpha_\ast$,
while the stars of the spectral class M6 have smallest $\alpha_\ast$.
The anisotropy of the convective turbulence decreases the values of $\alpha_\ast$
and causes a localization of the maximum value of the kinetic $\alpha$  effect
at the vicinity of the star surface and the equator.

\begin{table}
\caption[]{The coefficient $C_\ast$ for
different spectral classes and different rotation rates.}
\begin{flushleft}
\begin{tabular}{|c|c|c|r|}
\hline
spectral class & $\Omega_\odot$ & $10\Omega_\odot$ & $20\Omega_\odot$ \\
\hline
G2 &   0.970   &   0.933   & 0.982 \\
K2 &   0.877   &   0.874    & 0.883 \\
K4 &   0.824   &   0.858    & 0.854 \\
K7 &   0.855   &   0.815    & 0.793  \\
M2 &  0.643   &   0.687    & 0.680 \\
\hline
\end{tabular}\end{flushleft}
\end{table}

The maximum value of the kinetic $\alpha$ effect
can be estimated as  \citep{KRR95}
\begin{eqnarray}
\alpha_\ast = C_\ast (\Omega_\ast \, \eta_{_{T}})^{1/2} ,
\label{ALT1}
\end{eqnarray}
where the coefficient $C_\ast$ is given in Table~1 for
different spectral classes and different rotation rates.
It follows from Table~1 that the coefficient $C_\ast$ is weakly dependent on the rotation rates.
To derive equation~(\ref{ALT1}), we use a spatial distribution of the kinetic  $\alpha$
effect. In particular,  the kinetic  $\alpha$
effect is $ \alpha \simeq \ell(r) \Omega_\ast$ for $\ell(r) \Omega_\ast / u_c (r) \ll 1,$
and $\alpha \simeq u_c(r)$ for $\ell(r) \Omega_\ast / u_c (r) \gg 1 $
(Zeldovich et al. 1983). The kinetic  $\alpha$ effect reaches a maximum at the depth $r
= r_m$ determined by the condition $\ell_m (r_m)
= u_c(r_m) / \Omega_\ast$. The turbulent magnetic
diffusivity is $\eta_{_{T}} \simeq \ell_m (r_m) u_c (r_m)$. Therefore,
$\ell_m (r_m)\simeq ( \eta_{_{T}} /\Omega_\ast )^{1/2}$. The maximal
value of the kinetic $ \alpha$ effect, $\alpha_\ast $, is given by
$\alpha_\ast \simeq u_c (r_m) \simeq \eta_{_{T}} / \ell_m (r_m) \simeq
( \eta_{_{T}} \Omega_\ast )^{1/2}$.

In Figs.~\ref{Fig6}--\ref{Fig7} we show the radial profiles $\alpha_{\rm norm} =\alpha/(\Omega_\odot\, \eta_{_{T}})^{1/2}$
of the kinetic $\alpha$ effect normalized by $(\Omega_\odot\, \eta_{_{T}})^{1/2}$.
This normalization and anisotropy of the convective turbulence cause the curves of the different spectral classes
 $\alpha_{\rm norm}$ almost collapse to each other.
 This indicates that the estimate~(\ref{ALT1}) is enough good.
 In the present study we will use this estimate to determine
 the dynamo number (see below).

\section{Mean-field $\alpha^2 \, \Omega$ dynamo}
\label{sect3}

Mean-field theories of solar, stellar and galactic dynamos have been developing during last 55 years
\citep[see, e.g., books by][]{M78,P79,KR80,ZRS83,RSS88,RHK13,MD19,RI21,SS21}.
In the present study, we show that magnetic field generation in
fast rotating stars of the spectral classes: from M6 to G0 can be described
by the axisymmetric mean-field $\alpha^2 \, \Omega$ dynamo,
where the $\alpha^2$ dynamo is modified by a weak differential rotation
with $R_\omega \ll R_{\alpha} R_{\alpha}^{\rm cr}$.
The axisymmetric large-scale magnetic field can be written as  $ \meanBB = \meanB_\varphi
{\bm e}_{\varphi} + \bec{\nabla} {\bf \times} (\meanAAA {\bm e}_{\varphi})$,
where $r, \theta, \varphi$ are the spherical coordinates and ${\bm e}_{\varphi}$ is the unit vector.
We consider the mean-field dynamo in a thin convective shell,
taking into account strong variation of the plasma density
in the radial direction (see below).
We neglect the curvature of the convective shell
and replace it by a flat slab.
Thus, the mean-field $\alpha^2 \, \Omega$ dynamo equations are given by:
\begin{eqnarray}
{\partial \meanB_\varphi \over \partial t} &=& \left[R_\alpha \, R_\omega \, \sin \theta {\partial \over \partial \theta}
- R_\alpha^2 \,  \left({\partial^2 \over \partial \theta^2} - \mu^2 \right)
 \right]\meanA
\nonumber\\
&& + \left({\partial^2 \over \partial \theta^2} - \mu^2 \right)\meanB_\varphi ,
\label{TM1}
\end{eqnarray}
\begin{eqnarray}
{\partial \meanA \over \partial t} &=& \alpha \meanB_\varphi + \left({\partial^2
\over \partial \theta^2} - \mu^2 \right) \meanA .
\label{TM2}
\end{eqnarray}
To take into account strong variation of the plasma density in the radial direction, we
average the dynamo equations over the depth of the convective zone and use the no-$r$ model.
In particular, the terms describing turbulent diffusion of the mean magnetic field
in the radial direction in equations~(\ref{TM1}) and~(\ref{TM2}) in the framework
of the no-$r$ model are given as $-\mu^2 \meanB_\varphi$
and $-\mu^2\meanA$ \citep{KKMR03,KSR16,SKR18}.
The differential rotation is characterised by parameter
$G =\partial \Omega / \partial r$, and the parameter $\mu$ is determined by the following equation:
$\int_{r_{\rm c}}^{1} (\partial^2 \meanB_\varphi / \partial r^2) \,dr = - (\mu^2/3) \meanB_\varphi$.

Equations~(\ref{TM1})--(\ref{TM2}) are written in dimensionless variables:
the coordinate $r$ is measured in the units of the star radius $R_\ast$, the time $t$
is measured in the units of turbulent magnetic diffusion time $R_\ast^2 / \eta_{_{T}}$;
the toroidal component, $\meanB_\varphi(t,r,\theta)$, of the mean magnetic field is
measured in the units of $B_\ast$, where
$B_\ast \equiv  \xi \,\, \meanB_{\rm eq}$, $\xi= \ell_0/\sqrt{2}R_\ast$ and $\meanB_{\rm eq} = u_0 \, \sqrt{4 \pi \meanrho_\ast}$.
The  magnetic potential, $\meanA(t,r,\theta)$, of the poloidal field is measured in the units of
$R_{\alpha} R_\ast B_\ast$, where $R_{\alpha} = \alpha_\ast R_\ast / \eta_{_{T}}$,
the fluid density $\meanrho(r,\theta)$ is measured in the units $\meanrho_\ast$,
the differential rotation $\delta\Omega$ is measured in units of the maximal value
of the angular velocity $\Omega$,
the $\alpha$ effect is measured in units of the maximum value of the
kinetic $ \alpha $ effect, $\alpha_\ast$;
the integral scale of the turbulent motions
$\ell_0$ and the characteristic turbulent velocity $u_0$ at the scale $\ell_0$ are measured in units of their
maximum values in the convective region.
The magnetic Reynolds number ${\rm Rm}=\ell_0 \, u_0/\eta$
is defined using the maximal values of the integral scale $\ell_0$ and
the characteristic turbulent velocity $u_0$,
and the turbulent magnetic diffusion coefficient is $\eta_{_{T}}=\ell_0 \, u_0 / 3$.
The dynamo number is defined as $D = R_\alpha
R_\omega$, where $R_\omega = (\delta \Omega) \, R_\ast^2 / \eta_{_{T}}$.

Equations~(\ref{TM1}) and~(\ref{TM2})
describe the dynamo waves propagating from the central
latitudes towards the equator when the dynamo number is negative.
The radius $r$ varies from $r_{\rm c}$ to $1$
inside the convective shell, so that, e.g., for stars of the spectral class G2
(the solar-like stars),
the value $\mu=3$ corresponds to a convective
zone with a thickness of about 1/3 of the radius.

\subsection{Kinematic $\alpha^2 \, \Omega$ dynamo}
\label{subsect3.1}

First, we consider a kinematic dynamo problem,
assuming for simplicity that the kinetic $\alpha$ effect
is a constant.
Note that the kinematic and weakly non-linear $\alpha^2 \, \Omega$ dynamos have been studied in a number of publications
[see, e.g., \cite{MNS96,GSK01,BSK05}, and references therein].
We seek a solution for equations~(\ref{TM1})--(\ref{TM2}) as a real part of the following functions:
\begin{eqnarray}
\overline{A}=A_0 \exp(\tilde \gamma t - {\rm i} \, k \, \theta) ,
\label{TM3}
\end{eqnarray}
\begin{eqnarray}
\overline{B}_\varphi = B_0 \exp(\tilde \gamma t - {\rm i} \, k \, \theta) ,
\label{TM4}
\end{eqnarray}
where $\tilde \gamma=\gamma + {\rm i} \, \omega$.
Equations~(\ref{TM1})--(\ref{TM4}) yield the growth rate of the dynamo instability and the frequency of the dynamo waves as
\begin{eqnarray}
\gamma &=& {R_{\alpha} R_{\alpha}^{\rm cr} \over \sqrt{2}} \left[\left[1 + \left({\zeta R_\omega \over R_{\alpha} R_{\alpha}^{\rm cr}}\right)^2\right]^{1/2} + 1 \right]^{1/2}
- \left(R_{\alpha}^{\rm cr}\right)^2 ,
\nonumber\\
\label{TM5}
\end{eqnarray}
\begin{eqnarray}
\omega = - {\rm sgn}(R_\omega) \, {R_{\alpha} R_{\alpha}^{\rm cr} \over \sqrt{2}} \left[\left[1 + \left({\zeta R_\omega \over R_{\alpha} R_{\alpha}^{\rm cr}}\right)^2\right]^{1/2} - 1 \right]^{1/2}  ,
\label{TM6}
\end{eqnarray}
where $\zeta^2=1 - \left(\mu/R_{\alpha}^{\rm cr}\right)^2$.
Here we took into account that
$(x + {\rm i}y)^{1/2}= \pm (X + {\rm i}Y)$, where
$X = 2^{-1/2} \, [(x^2+y^2)^{1/2} + x]^{1/2}$ and $Y = {\rm sgn}(y) \, 2^{-1/2} \, [(x^2+y^2)^{1/2} - x]^{1/2}$.
Here the threshold $R_{\alpha}^{\rm cr}$ for the mean-field dynamo instability, defined by the conditions
$\gamma=0$ and $R_\omega=0$, is given by $R_{\alpha}^{\rm cr}=(k^2 + \mu^2)^{1/2}$.

Equations~(\ref{TM1})--(\ref{TM4}) allow one to determine the squared amplitude ratio $|A_0 / B_0|^2$ as
\begin{eqnarray}
\left|{A_0 \over B_0}\right|^2 =  \left(R_{\alpha} R_{\alpha}^{\rm cr} \right)^{-2} \,
\left[1 + \left({\zeta R_\omega \over R_{\alpha} R_{\alpha}^{\rm cr}}\right)^2\right]^{-1/2} ,
\label{TM11}
\end{eqnarray}
and the phase shift $\delta$ between the toroidal field $\overline{B}_\varphi$ and the magnetic vector potential $\overline{A}$ is given by
\begin{eqnarray}
\sin(2\delta) =  - \zeta R_\omega \,  \left[\left(R_{\alpha} R_{\alpha}^{\rm cr} \right)^{2} + \zeta^2 R_\omega^2\right]^{-1/2} .
\label{TM14}
\end{eqnarray}
Equation~(\ref{TM11}) yields the energy ratio of poloidal $\overline{B}_{\rm pol} =R_{\alpha} R_{\alpha}^{\rm cr} \, \overline{A}$ and toroidal $\overline{B}_\varphi$ mean magnetic field components as
\begin{eqnarray}
{\meanB_{\rm pol}^2 \over \meanB_\varphi^2} =  \left[1 + \left({\zeta R_\omega \over R_{\alpha} R_{\alpha}^{\rm cr}}\right)^2\right]^{-1/2} .
\label{TM12}
\end{eqnarray}

Asymptotic formulas for the growth rate of the dynamo instability and the frequency of the dynamo waves for a weak differential rotation, $\zeta R_\omega \ll R_{\alpha} R_{\alpha}^{\rm cr}$,  are given by
\begin{eqnarray}
\gamma = R_{\alpha} R_{\alpha}^{\rm cr} \left[1 + {1 \over 8} \left({\zeta R_\omega \over R_{\alpha} R_{\alpha}^{\rm cr}}\right)^2\right] - \left(R_{\alpha}^{\rm cr}\right)^2 ,
\label{TM7}
\end{eqnarray}
\begin{eqnarray}
\omega =- {\zeta R_\omega \over \sqrt{2}} .
\label{TM8}
\end{eqnarray}
In this case, the mean-field $\alpha^2$ dynamo is slightly modified by a weak differential rotation, and
the phase shift between the fields $\overline{B}_\varphi$ and $\overline{B}_\theta$ vanishes,
while $\overline{B}_{\rm pol}/  \overline{B}_\varphi \sim 1$ [see equations~(\ref{TM14})--(\ref{TM12})].
The period of the dynamo wave is $T_{\rm dyn}=(2 \pi /\omega) \, (R_\ast^2 / \eta_{_{T}})$,
where $\omega$ is the non-dimensional frequency of the dynamo wave given by equation~(\ref{TM8}).
In the present study, we show that this case corresponds to fast rotating stars
of the spectral class from M6 to G0. Since in this case $\overline{B}_{\rm pol} \sim \overline{B}_\varphi$,
the star spots can be formed for any latitude.

In the opposite case, for a strong differential rotation, $\zeta R_\omega \gg R_{\alpha} R_{\alpha}^{\rm cr}$, the growth rate of the dynamo instability and the frequency of the dynamo waves are given by
\begin{eqnarray}
\gamma = \left[ {1 \over 2} \,  \zeta \, R_{\alpha}^{\rm cr} \, R_{\alpha} |R_\omega| \right]^{1/2} - \left(R_{\alpha}^{\rm cr}\right)^2 ,
\label{TM9}
\end{eqnarray}
\begin{eqnarray}
\omega = - {\rm sgn}(R_\omega) \left[ {1 \over 2} \,  \zeta \, R_{\alpha}^{\rm cr} \, R_{\alpha} |R_\omega| \right]^{1/2} .
\label{TM10}
\end{eqnarray}
In this case the mean-field $\alpha\Omega$ dynamo is slightly modified by a weak $\alpha^2$ effect,
and the phase shift between the fields $\overline{B}_\varphi$ and $\overline{B}_\theta$ tends to $- \pi/4$,
while $\overline{B}_{\rm pol}/  \overline{B}_\varphi \ll 1$ [see equations~(\ref{TM14})--(\ref{TM12})].
This case corresponds to the solar dynamo.
The necessary condition for the dynamo ($\gamma>0$) is

(a) when $R_{\alpha}/R_{\alpha}^{\rm cr} < \sqrt{2}$, the mean-field $\alpha^2 \, \Omega$ dynamo
is excited when
\begin{eqnarray}
D > {2 \over \zeta} \, \left(R_{\alpha}^{\rm cr}\right)^3 ;
\label{TL1}
\end{eqnarray}

(b) when $R_{\alpha}/R_{\alpha}^{\rm cr} > \sqrt{2}$, the mean-field $\alpha^2 \, \Omega$ dynamo
is excited for any differential rotation, $R_\omega$.

\subsection{Algebraic and dynamic non-linearities}
\label{subsect3.2}

Now we discuss the algebraic and dynamic non-linearities in the non-linear dynamo model.
The total $\alpha$ effect is the sum of the kinetic and magnetic $\alpha$ effects,
\begin{eqnarray}
\alpha = \chi_{_{\rm K}} \Phi_{_{\rm K}}(\meanB) + \sigma_\rho \chi_{_{\rm M}} \Phi_{_{\rm M}}(\meanB) ,
\label{AL1}
\end{eqnarray}
where $\chi_{_{\rm K}} = - (\tau_0 /3) \, \langle\bec{u}\cdot(\bec{\nabla}
{\bf \times} \bec{u})\rangle$ is proportional to the kinetic helicity $\langle{\bm u} \cdot (\bec{\nabla} {\bf \times} {\bm u})\rangle$
and  $\chi_{_{\rm M}} = (\tau_0 / 12 \pi \meanrho)\,\langle{\bm b} \cdot (\bec{\nabla} {\bf \times} {\bm b})\rangle$ is proportional to the current helicity $\langle{\bm b} \cdot (\bec{\nabla} {\bf \times} {\bm b})\rangle$
 \citep{FPL75,PFL76}.
Here $\tau_0$ is the correlation time of the turbulent velocity field,
${\bm u}$ and ${\bm b}$ are velocity and magnetic fluctuations,
and $\sigma_\rho = \int_{r_{\rm c}}^{1} (\meanrho(r)/ \meanrho_\ast)^{-1} \, dr$  \citep{KSR16,SKR18}.

The quenching functions $\Phi_{_{\rm K}}(\meanB)$ and $\Phi_{_{\rm M}}(\meanB)$ in
equation for the total $\alpha$ effect are given by \citep{RK00,RK01,RK04,RK06},
\begin{eqnarray}
\Phi_{_{\rm K}}(\meanB) = {1 \over 7} \left[4 \Phi_{_{\rm M}}(\meanB) + 3 \Phi_{_{\rm B}}(\meanB)\right] ,
\label{AL2}
\end{eqnarray}
and \citep{FBC99}
\begin{eqnarray}
\Phi_{_{\rm M}}(\meanB) = {3\over 8\beta^2} \, \left[1 - {\arctan (\sqrt{8} \beta) \over
\sqrt{8} \beta} \right] ,
\label{AL3}
\end{eqnarray}
where $\beta=\meanB/\meanB_{\rm eq}$
\begin{eqnarray}
\Phi_{_{\rm B}}(\meanB) = 1 - 16 \beta^{2} + 128 \beta^{4} \ln \left[1 + (8\beta^2)^{-1}\right],
\label{AL4}
\end{eqnarray}
and $\chi_{_{\rm K}}$ and $\chi_{_{\rm M}}$ are measured in units of maximal
value of the $\alpha$-effect, $\alpha_\ast$.
The function $\Phi_{_{\rm K}}$ describes the algebraic quenching of the
kinetic $\alpha$ effect that is caused by the feedback effects of the mean magnetic field
on plasma motions.
The densities of the kinetic and current helicities, and quenching functions are
associated with a middle part of the convective zone.
The parameter $\sigma_\rho > 1$ is a free parameter.

The magnetic $\alpha$ effect, $\alpha_{_{\rm M}}$,  is based on two
non-linearities: the algebraic non-linearity due to the feedback effects of the mean magnetic field
on plasma motions, that is described by the quenching function
$\Phi_{_{\rm M}}(\meanB)$, and the dynamic non-linearity, characterised by
the function $\chi_{_{\rm M}}(\meanBB)$ that is determined by a dynamical
equation \citep{KR82,GD94,KR99,KRR95,KMR00,KMR02,KKMR03,KMR03,BF00,BS05,ZMKR06,ZMKR12}.
In particular, the total magnetic helicity, $\int (H_{\rm M} + H_{\rm m}) \,dr^3$, is conserved for very small microscopic magnetic diffusivity $\eta$, where $H_{\rm M}=\overline{\bm A} {\bf \cdot} \overline{\bm B}$
is the magnetic helicity density of the large-scale field $\overline{\bm B}=\bec{\nabla} {\bf \times} \overline{\bm A}$
with $\overline{\bm A}$ being the mean magnetic vector potential,
$H_{\rm m} =\langle {\bm a} {\bf \cdot} {\bm b} \rangle$ is the magnetic helicity density of the small-scale field ${\bm b}=\bec{\nabla} {\bf \times} {\bm a}$ with ${\bm a}$ being fluctuations of magnetic vector potential.

When the mean-field dynamo amplifies the large-scale magnetic field,
the magnetic helicity density $H_{\rm M}$ of the large-scale field grows in time.
Since the total magnetic helicity $\int (H_{\rm M} + H_{\rm m}) \,dr^3$
is conserved, the magnetic helicity density $H_{\rm m}$ of the small-scale field changes during
the dynamo action, and its evolution is determined by
the non-dimensional dynamical equation reads \citep{KSR16,SKR18}:
\begin{eqnarray}
&& {\partial \chi_{\rm c} \over \partial t} + \left(\tau_\chi^{-1} + \kappa_{_{T}}
\mu^2\right)\chi_{\rm c} = 2\left({\partial
\meanAAA \over \partial \theta} {\partial \meanB_\varphi \over \partial \theta} + \mu^2 \meanAAA \, \meanB_\varphi\right)
\nonumber\\
&& \quad -  \alpha \, \meanB^2 - {\partial \over \partial \theta} \left(\meanB_\varphi {\partial \meanAAA \over \partial \theta} - \kappa_{_{T}} {\partial \chi^c
\over \partial \theta} \right) ,
\label{TS1}
\end{eqnarray}
where ${\bm F}_\chi = -\kappa_{_{T}} \bec{\nabla} \chi_{\rm c}$
is the turbulent diffusion flux of the magnetic helicity density of small-scale field and
$\kappa_{_{T}}$ is the coefficient of the turbulent diffusion of the magnetic helicity.
Dynamics of magnetic helicity of small-scale field is a crucial mechanism
in a non-linear dynamo saturation where turbulent magnetic helicity fluxes allow to avoid catastrophic
quenching of the $\alpha$ effect.
Recently, turbulent fluxes of magnetic helicity density of small-scale magnetic field have been rigorously derived
by \cite{KR22,GS23}.

In equation~(\ref{TS1}), the time $\tau_\chi = \ell^2 / \eta$ is the relaxation time of magnetic helicity.
The average value of $\tau_\chi^{-1}$ is given by
\begin{eqnarray}
\tau_\chi^{-1} = H_\ast^{-1} \int_{r_{\rm c}}^{1} \tilde \tau_\chi^{-1}(r) \,d r \sim {H_\ell \, R_\ast^2 \,
\eta \over H_\ast \, \ell^2 \, \eta_{_{T}}} ,
\label{TS2}
\end{eqnarray}
where $H_\ast$ is the depth of the convective zone,
$H_\ell$ is the characteristic scale of variations $\ell_0$, and
$\tilde \tau_\chi(r) = (\eta_{_{T}} / R_\ast^{2}) (\ell_0^2 / \eta)$ is the non-dimensional
relaxation time of the density of the magnetic helicity. The values
$H_\ell , \, \eta, \, \ell_0$ in equation~(\ref{TS2}) are associated
with the upper part of the convective zone.
The mean magnetic field is given by
\begin{eqnarray}
\meanB^2 = \meanB_\varphi^2 + R_\alpha^2 \left[\mu^2 \meanAAA^2 + \left( {\partial \meanAAA \over \partial \theta}\right)^2\right] .
\label{TS3}
\end{eqnarray}

\subsection{Mean-field numerical simulations of the $\alpha^2\Omega$ dynamo}
\label{subsect3.3}

We perform the mean-field numerical simulations of the $\alpha^2\Omega$ dynamo by
solving numerically equations~(\ref{TM1}), (\ref{TM2})  and (\ref{TS1}).
We use MATLAB code, which solves initial-boundary value problems for systems of parabolic and elliptic partial-differential equations
that employs a second-order explicit finite differences in space. We use the spatial resolution of $203$ mesh points in co-latitude $\theta$
(this odd number provides mesh intervals below 1 degree).
The time grid in simulations varied between $6 \times 10^5$ and $18 \times 10^5$ time instants for a different set of initial parameters
due to long transitional processes.

For numerical simulations, we use the standard profile of the kinetic $\alpha$ effect:
$\alpha(\theta)=\alpha_0 \sin^3 \theta \cos \theta$.
We use the following initial conditions: $\meanB_\phi(t=0,\theta)=S_1 \sin\theta + S_2 \sin(2\theta)$ and $\meanAAA(t=0,\theta)=0$
corresponding to a combination of the dipole and quadropole type solutions.
The parameters of the numerical simulation are as follows: $G=1$, $\xi=0.1$, $\kappa_{_{T}}=0.1$, $T=6.3$, $S_1=0.051$, $S_2=0.95$ for different $\mu$, $R_\alpha$ and $R_\omega$.
These parameters and initial conditions have been used by us for modelling of the solar activity
by the axisymmetric mean-field $\alpha \, \Omega$ dynamo \citep{KSR16,KSR20,SKR18}, where mechanism of the sunspot formation
related to negative effective magnetic pressure instability have been taken into account \citep{KRR89,KRR90,KMR96,KR94,RK07,BKR11,BRK16,WKR13,WKR16}.

\begin{figure}
\centering
\includegraphics[width=8.5cm]{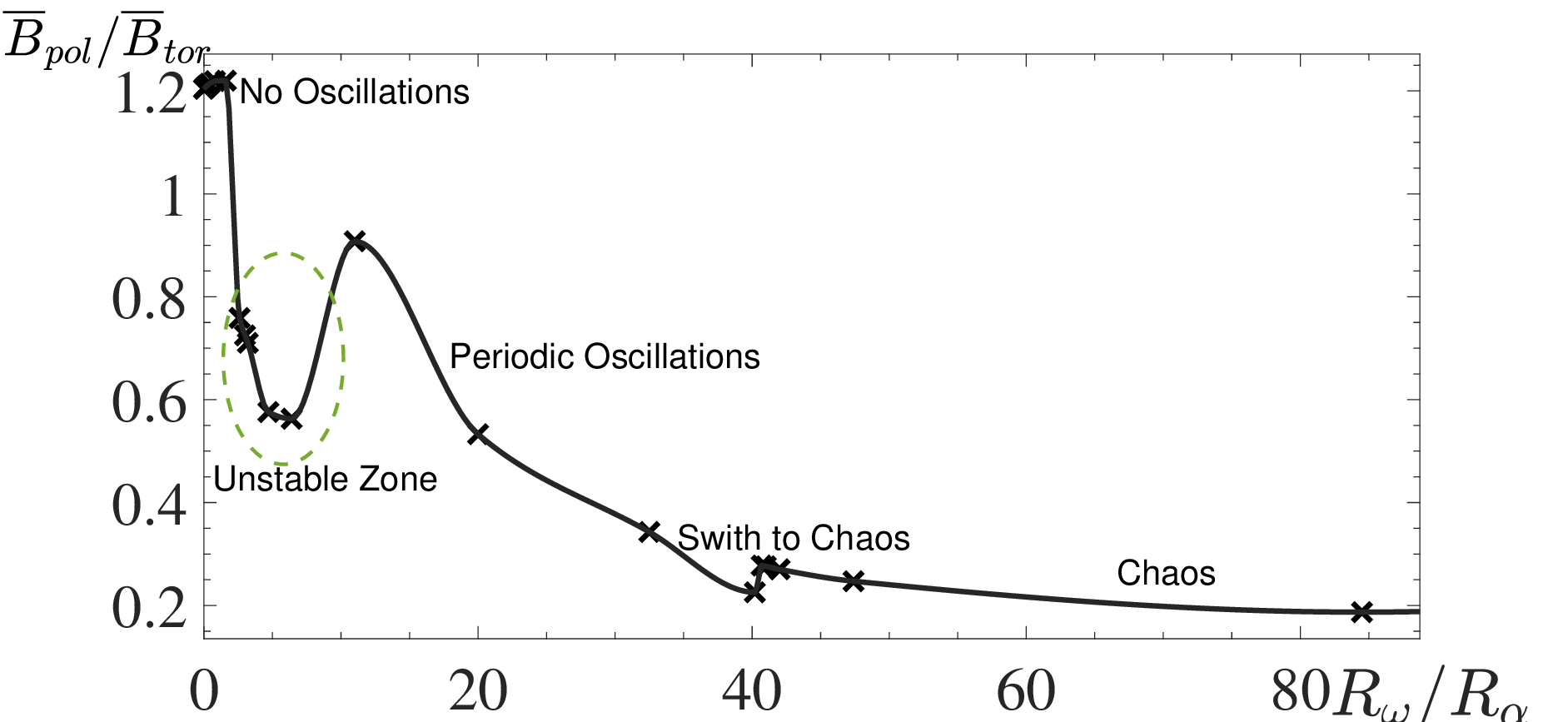}
\caption{\label{Fig8}
The ratio of the maximum values of the poloidal to toroidal mean magnetic fields $\meanB_{\rm pol}/\meanB_{\rm tor}$ versus $R_{\omega}/R_{\alpha}$  obtained from numerical simulations of $\alpha^2\Omega$ mean-field dynamo.
}
\end{figure}

\begin{figure}
\centering
\includegraphics[width=8.5cm]{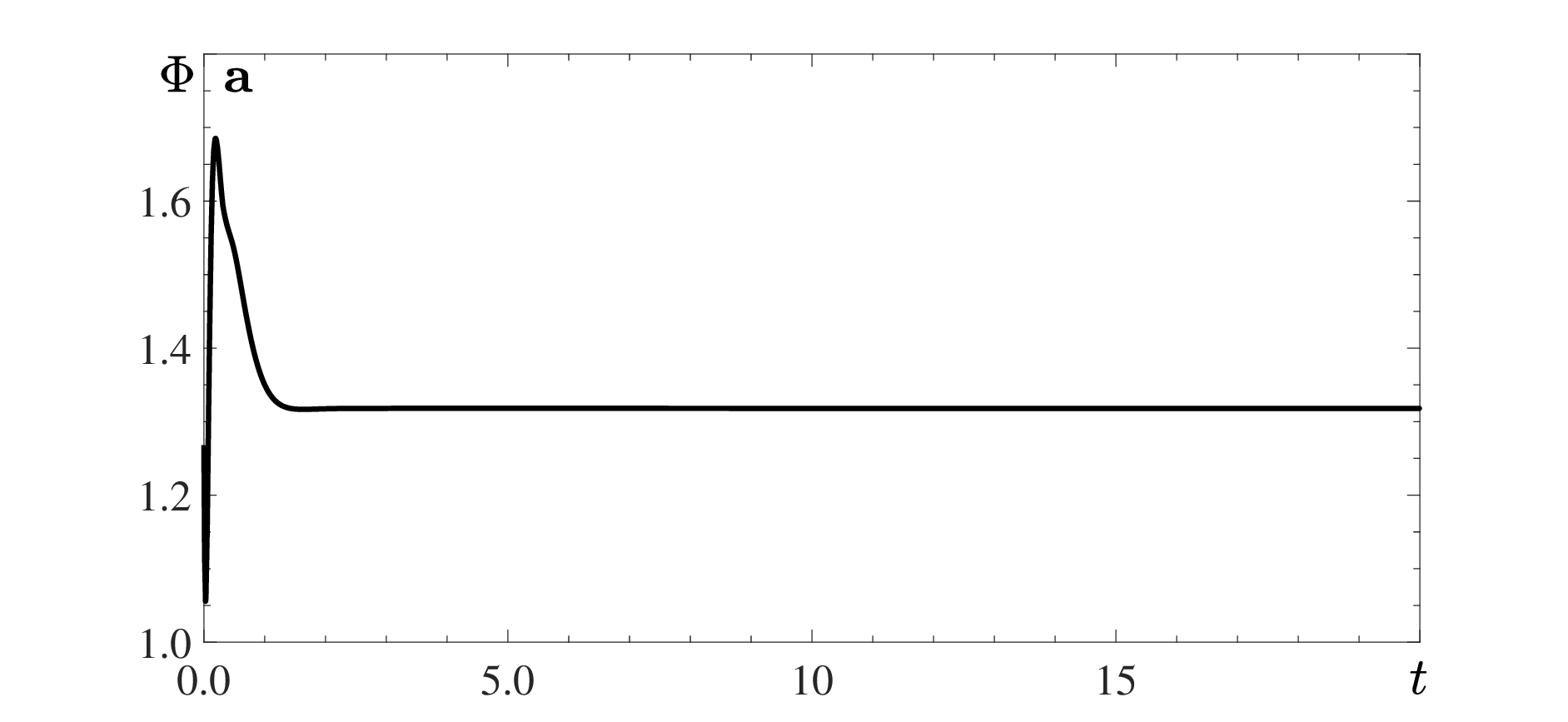}
\includegraphics[width=8.5cm]{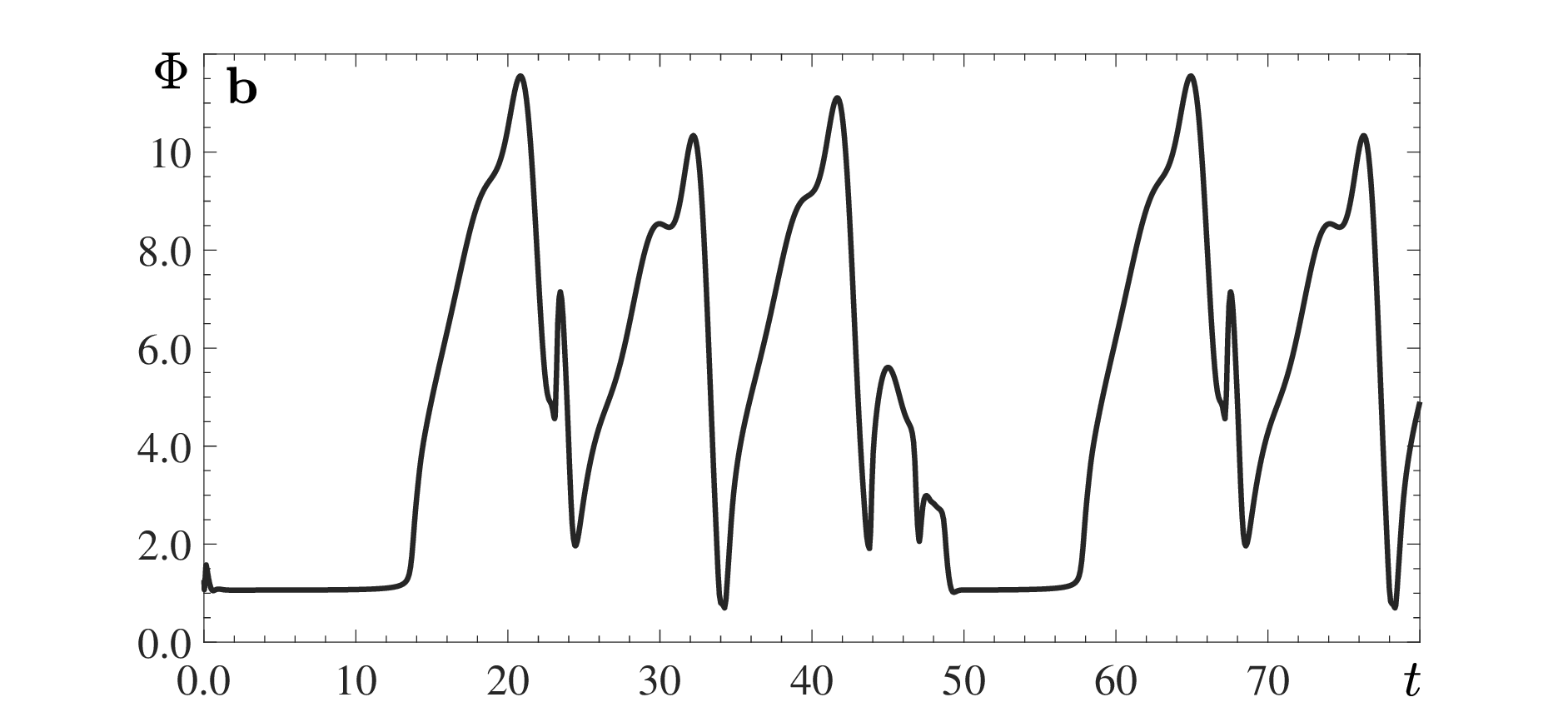}
\includegraphics[width=8.5cm]{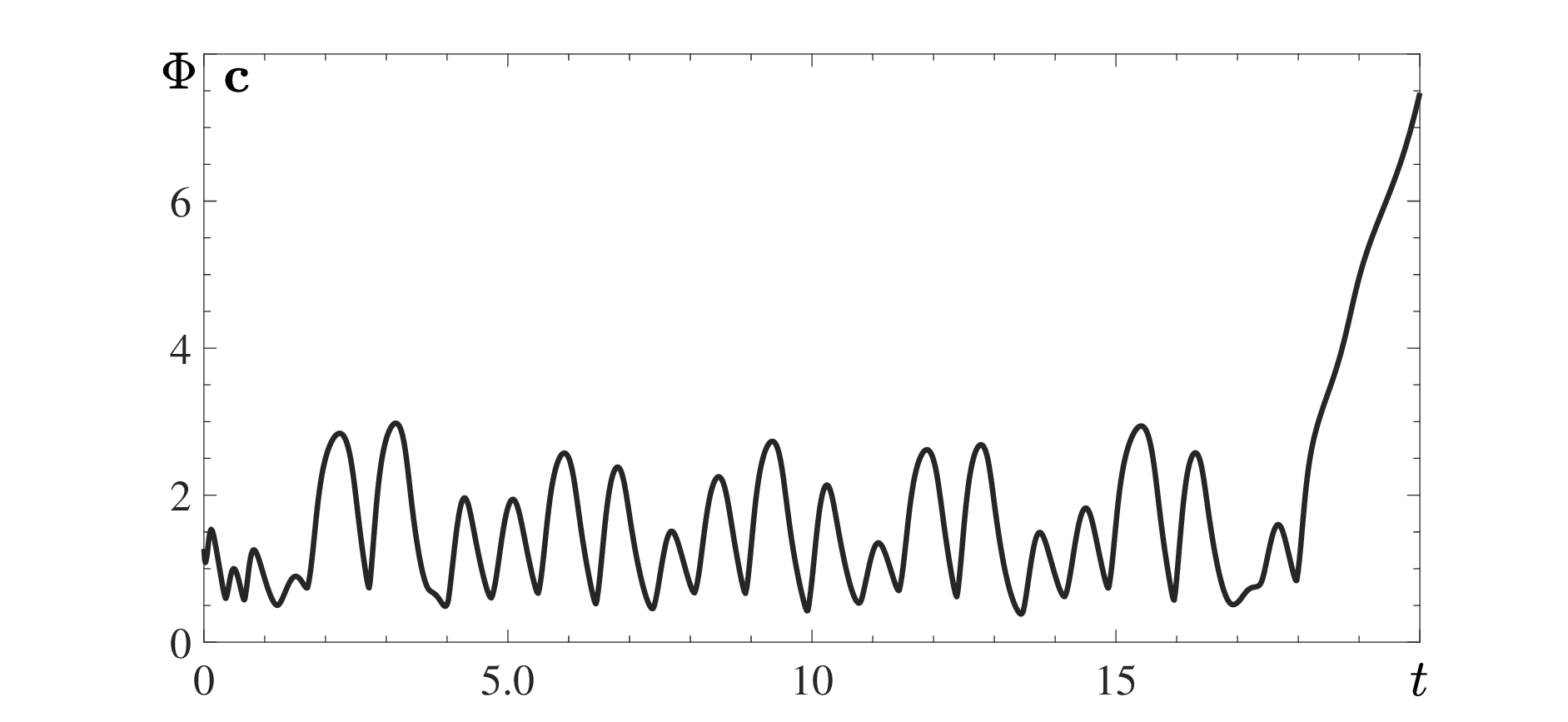}
\includegraphics[width=8.5cm]{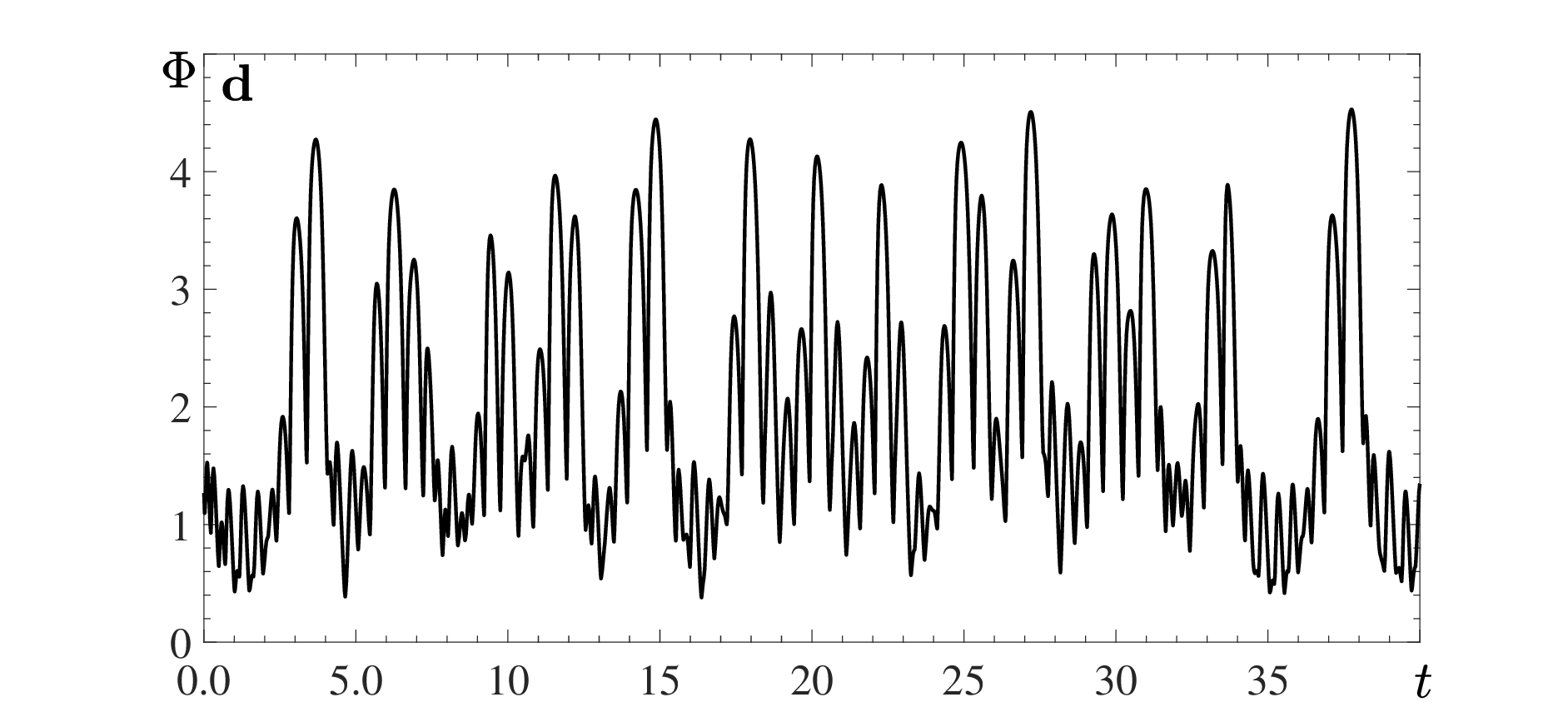}
\caption{\label{Fig9}
The time evolution of the flux of the toroidal mean magnetic field $\Phi = \int |\meanB_\varphi| \,d\sigma$
obtained from numerical simulations of $\alpha^2\Omega$ mean-field dynamo
at different values of $R_{\omega}/R_{\alpha}=1.6$ ({\bf a}); 3.2  ({\bf b}); 4.7   ({\bf c}) and 6.4  ({\bf d}).
The time is normalized by 122.2 years, and the flux $\Phi$ is
normalized by the magnetic field of 300 G.
}
\end{figure}

\begin{figure}
\centering
\includegraphics[width=8.5cm]{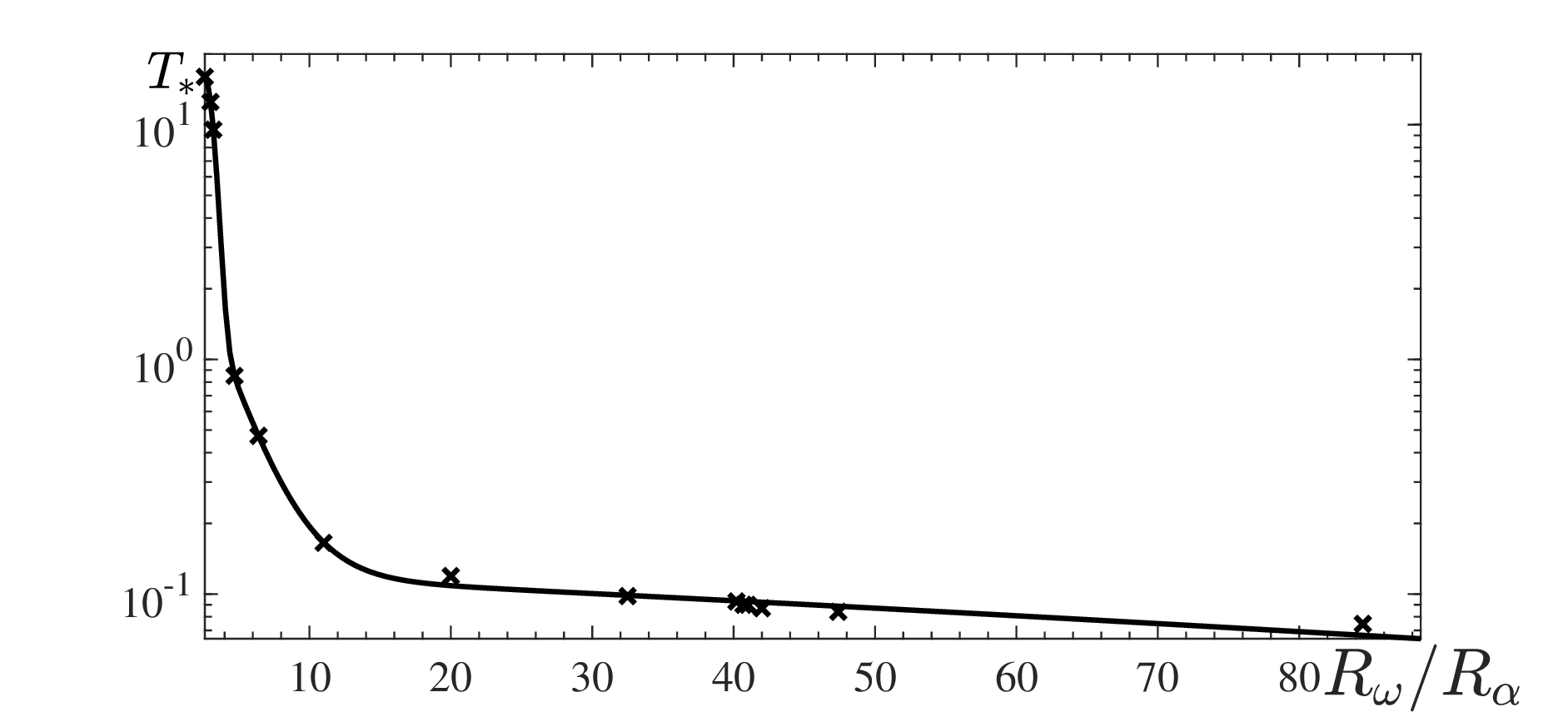}
\caption{\label{Fig10}
The period $T_\ast$ of the stellar magnetic cycles
normalized by 122.2 years versus $R_\omega/R_\alpha$
obtained from numerical simulations of the non-linear $\alpha^2\Omega$ mean-field dynamo.
}
\end{figure}

First we perform numerical simulations of the $\alpha^2\Omega$ mean-field dynamo at $R_{\alpha}=10$.
This value of the parameter $R_{\alpha}$ corresponds to the kinetic $\alpha$ effect arising in rotating convective turbulence
with the rotating frequency that is in 25 times larger than that for the Sun.
In Fig.~\ref{Fig8} we plot the ratio of the maximum values of the poloidal to toroidal mean magnetic fields
$\meanB_{\rm pol}/\meanB_{\, \rm tor}$ versus $R_{\omega}/R_{\alpha}$.
Depending on the ratio $R_{\omega}/R_{\alpha}$, there are ranges
of the aperiodic behavior, the quasi-periodic oscillations of the mean magnetic field
and the chaotic behaviour.
This is seen in Fig.~\ref{Fig9}, where we show the time evolution of the
flux of the toroidal mean magnetic field $\Phi = \int |\meanB_\varphi| \,d\sigma$
obtained from numerical simulations of the $\alpha^2\Omega$ mean-field dynamo
for different values of $R_{\omega}/R_{\alpha}=$1.6; 3.2; 4.7 and 6.4.
The time is normalized by 122.2 years, and the flux $\Phi$ is
normalized by the magnetic field of 300 G.

Note that 122.2 years corresponds to the turbulent diffusion time $R^2_\ast/\eta_{_{T}}$
with $R_\ast=R_\odot$ and $\eta_{_{T}}= 1.4 \times 10^{12}$ cm$^2$/s.
Here we take into account that the integral scale $\ell_0$
of the turbulent convection is smaller by a factor 5 --7  than the size of the coherent
structures (the large-scale circulations).
The latter is justified by the results of analytical study \citep{EKRZ02,EKRZ06,EGKR06} and
laboratory experiments \citep{BER09}.
This causes the mixing length used in the mixing length theory is about 5 --7 times larger than
the integral scale $\ell_0$ of the turbulent convection.
Correspondingly, the turbulent diffusion coefficients should be
5 --7 times smaller than those from the mixing length theory.

Increase of $R_{\omega}/R_{\alpha}$ causes decrease the periods $T_\ast$ of the stellar magnetic cycles.
This is seen in Fig.~\ref{Fig10}, where we show the periods $T_\ast$ of the stellar magnetic cycles
normalized by 122.2 years versus $R_{\omega}/R_{\alpha}$, which decreases from about $10^3$ years to 10 years
depending on the value of the differential rotation.
In chaotic regime there can be transition from one attractor with a short period
(of several tens years) to that of a larger period (of thousand years), see Fig.~\ref{Fig9}c.
For larger values of $R_{\omega}/R_{\alpha}$, the dynamo is similar to the
$\alpha\Omega$ mean-field dynamo, while
for small values of $R_{\omega}/R_{\alpha}$, the dynamo is similar to the
$\alpha^2$ mean-field dynamo.

We will show in the next section that for low-mass main sequences stars
rotating much faster than the Sun, the generated large-scale magnetic field
is caused by the mean-field $\alpha^2\Omega$ dynamo,
whereby the $\alpha^2$ dynamo is slightly modified by a weak differential rotation.
This means that $R_\omega \ll R_{\alpha} R_{\alpha}^{\rm cr}$.

\subsection{Mean-field numerical simulations of the $\alpha^2$ dynamo}
\label{subsect3.4}

\begin{figure}
\centering
\includegraphics[width=8.5cm]{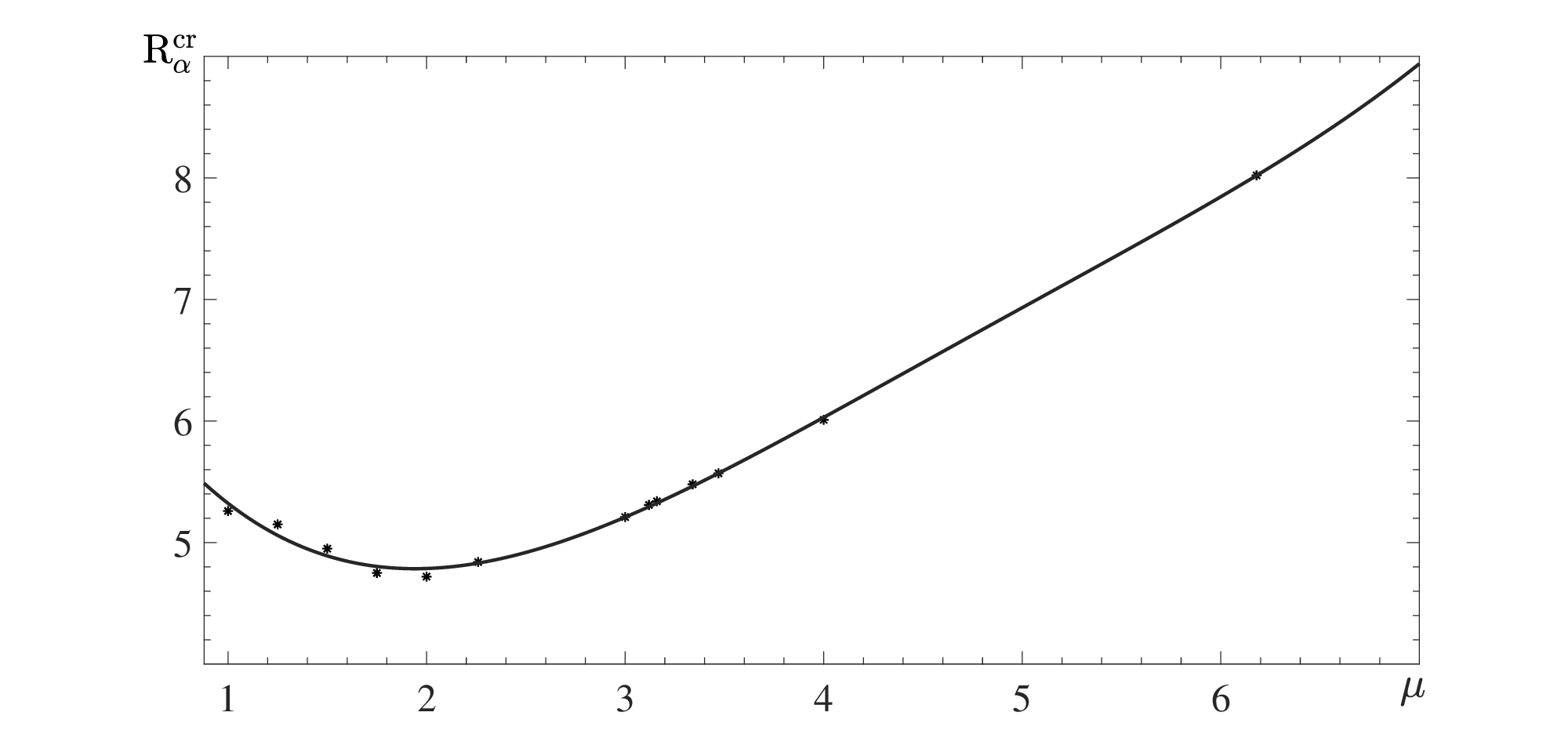}
\caption{\label{Fig11}
The threshold, $R_{\alpha}^{\rm cr}$, in generation of the large-scale magnetic field (snowflakes)
versus parameter $\mu$ obtained from numerical simulations of the non-linear $\alpha^2$ mean-field dynamo.
Here the solid line corresponds to the fitting curve.
}
\end{figure}

\begin{figure}
\centering
\includegraphics[width=8.5cm]{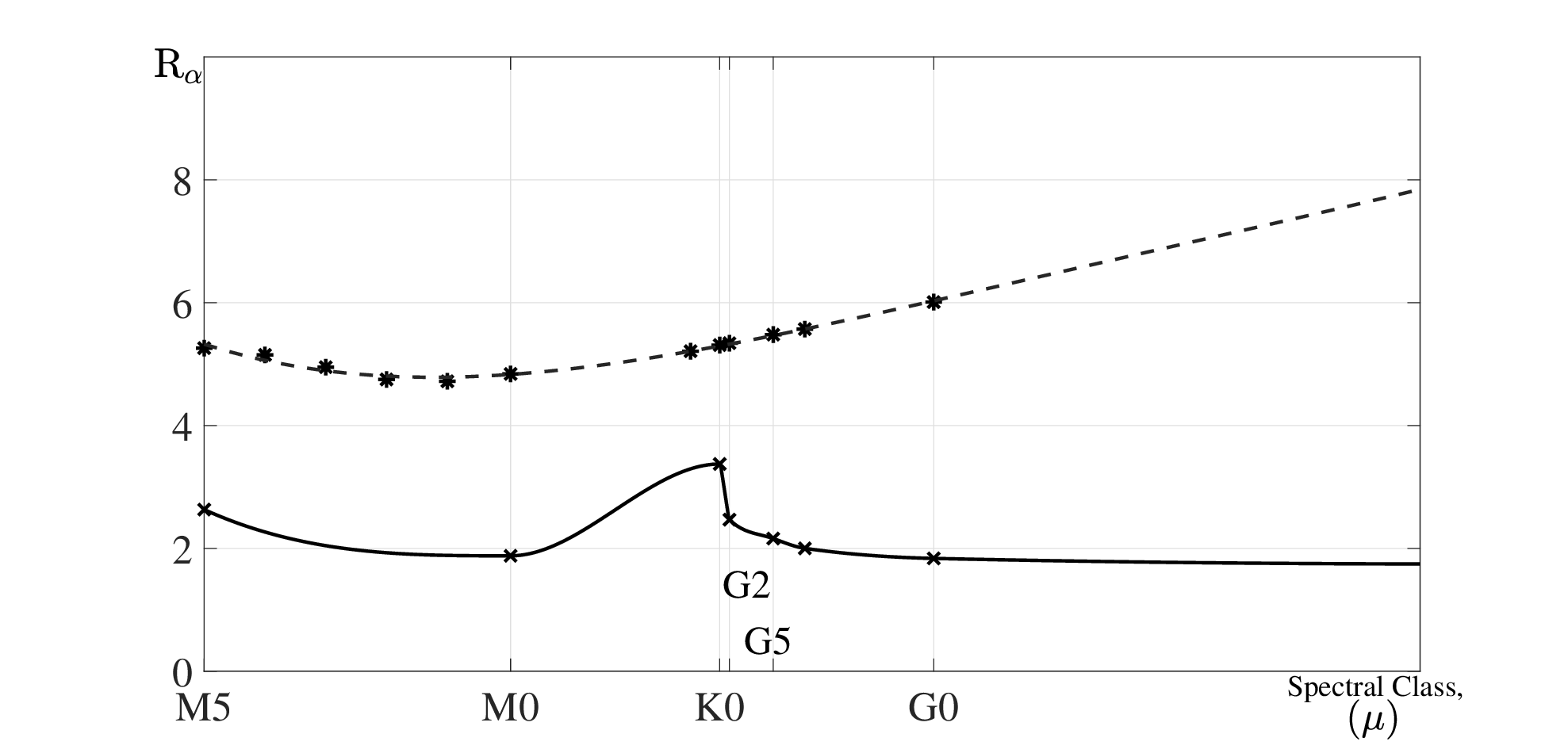}
\caption{\label{Fig12}
The threshold $R_{\alpha}^{\rm cr}$ (snowflakes, dashed line) in generation of the large-scale magnetic field
and the parameter $R_{\alpha}$ (crosses, solid line) versus the parameter $\mu$ obtained from numerical simulations of the non-linear $\alpha^2$ mean-field dynamo.
The parameter $R_{\alpha}= (\Omega_\odot /\eta_{_{T}})^{1/2} \, R_\ast $ is determined for the main sequence stars, where
the angular velocity coincides with the mean (averaged over the latitude) solar angular velocity $\Omega_\odot$.
}
\end{figure}

\begin{figure}
\centering
\includegraphics[width=8.5cm]{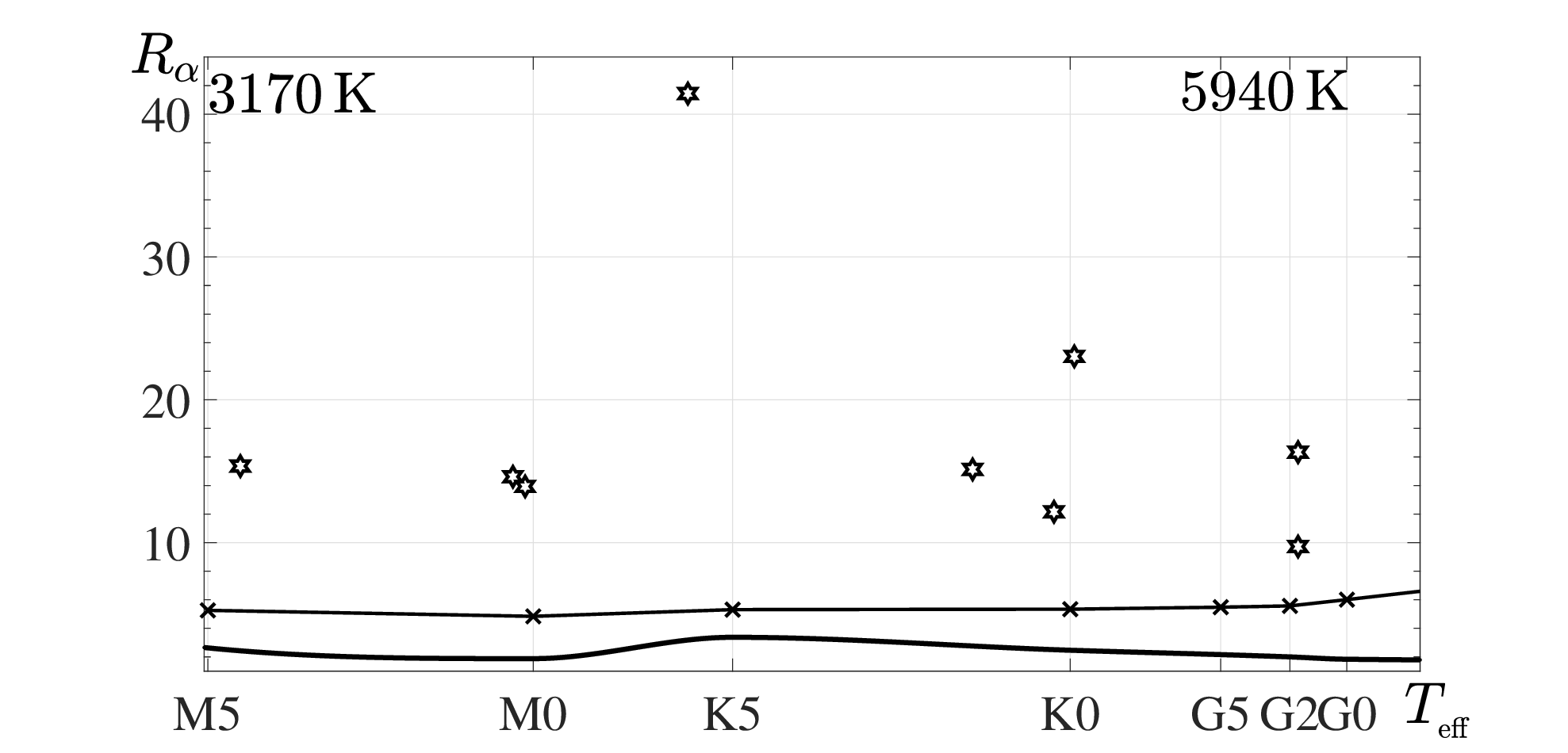}
\caption{\label{Fig13}
The threshold $R_{\alpha}^{\rm cr}$ (crosses, thin solid line) in generation of the large-scale magnetic field
and the parameter $R_{\alpha}=\left(\Omega_\odot /\eta_{_{T}}\right)^{1/2} \, R_\ast$ (solid line) versus the star effective temperature $T_{\rm eff}$
obtained from numerical simulations of the non-linear $\alpha^2$ mean-field dynamo.
The parameter $R_{\alpha}$ is calculated for the main sequence stars,
where the angular velocity coincides with the mean (averaged over the latitude) solar angular velocity $\Omega_\odot$.
The parameter $\tilde R_{\alpha}$ (stars) is also estimated for real main sequence stars, where we use equation~(\ref{ALT1}),
the rotating rates (see Gershberg et al. 2021) and turbulent magnetic diffusion  coefficients for the stars of these spectral classes.
}
\end{figure}

We also perform numerical simulations of the $\alpha^2$ mean-field dynamo with $R_\omega=0$.
First, we plot the threshold, $R_{\alpha}^{\rm cr}$, required for the generation of the large-scale magnetic field
versus
\begin{itemize}
\item{
parameter $\mu$ (Fig.~\ref{Fig11}),}
\item{
the spectral class (Fig.~\ref{Fig12}) and}
\item{
the stellar effective temperature $T_{\rm eff}$ (Fig.~\ref{Fig13}),}
\end{itemize}
obtained from numerical simulations.
For $\mu \geq 3$, the function $R_{\alpha}^{\rm cr}(\mu)$ is closed to the linear one (see Fig.~\ref{Fig11}).
Indeed, our asymptotic analysis for a constant kinetic $\alpha$ effect shows that $R_{\alpha}^{\rm cr}
=(k^2 + \mu^2)^{1/2}$. This implies that when $k^2 \ll \mu^2$,
we obtain that $R_{\alpha}^{\rm cr} \sim \mu$.

In Fig.~\ref{Fig12}, in addition to the threshold $R_{\alpha}^{\rm cr}$ versus the stellar spectral class, we also plot the parameter $R_{\alpha}\equiv  \alpha_\ast R_\ast / \eta_{_{T}}= (\Omega_\odot /\eta_{_{T}})^{1/2} \, R_\ast$ based on the solar
rotation rate [see equation~(\ref{ALT1})].
As follows from Fig.~\ref{Fig12}, the parameter  $R_{\alpha}$ is in several times less than
the threshold $R_{\alpha}^{\rm cr}$ required for the generation of the large-scale magnetic field.
This implies that the pure $\alpha^2$ dynamo with the kinetic $\alpha$ effect $\alpha=(\Omega_\odot \eta_{_{T}})^{1/2}$
based on the solar rotation rate
cannot explain the generation of large-scale magnetic field in the main sequences stars.
To describe correctly the magnetic field generation in the main sequences stars in the framework of the $\alpha^2$ dynamo,
one need to increase the stellar rotation rate by one order of magnitude to obtain the required value  of the $\alpha$ effect.
That is why we consider stars rotating much faster than the Sun.

In Fig.~\ref{Fig13}, we show the threshold $R_{\alpha}^{\rm cr}$ (crosses, thin solid line) in generation of the large-scale magnetic field
and the parameter $R_{\alpha}=\left(\Omega_\odot /\eta_{_{T}}\right)^{1/2} \, R_\ast$ (thick solid line) versus the star effective temperature $T_{\rm eff}$ obtained from numerical simulations of the non-linear $\alpha^2$ mean-field dynamo.
This parameter $R_{\alpha}$ is calculated for the main sequence stars,
where the angular velocity coincides with the mean (averaged over the latitude) solar angular velocity $\Omega_\odot$.
In addition, we also show the parameter $\tilde R_{\alpha}$ (shown as stars) that is estimated
for real main sequence stars, where we use equation~(\ref{ALT1}),
the rotating rates   (see Gershberg et al. 2021) and turbulent magnetic diffusion coefficients for the stars of these spectral classes.
Figure~\ref{Fig13} demonstrates that the parameter $\tilde R_{\alpha}$ for the observed stars is in several times
larger than the threshold $R_{\alpha}^{\rm cr}$ required for the generation of the large-scale magnetic field by pure $\alpha^2$ dynamo.
This shows that the pure $\alpha^2$ dynamo can describe the generation of large-scale magnetic field for these stars.
However, some observed features (appearance of star spots in the polar regions, long period of cyclic behaviour, etc.)
for the main sequence fast rotating stars require presence of small differential rotation.

\begin{figure}
\centering
\includegraphics[width=8.5cm]{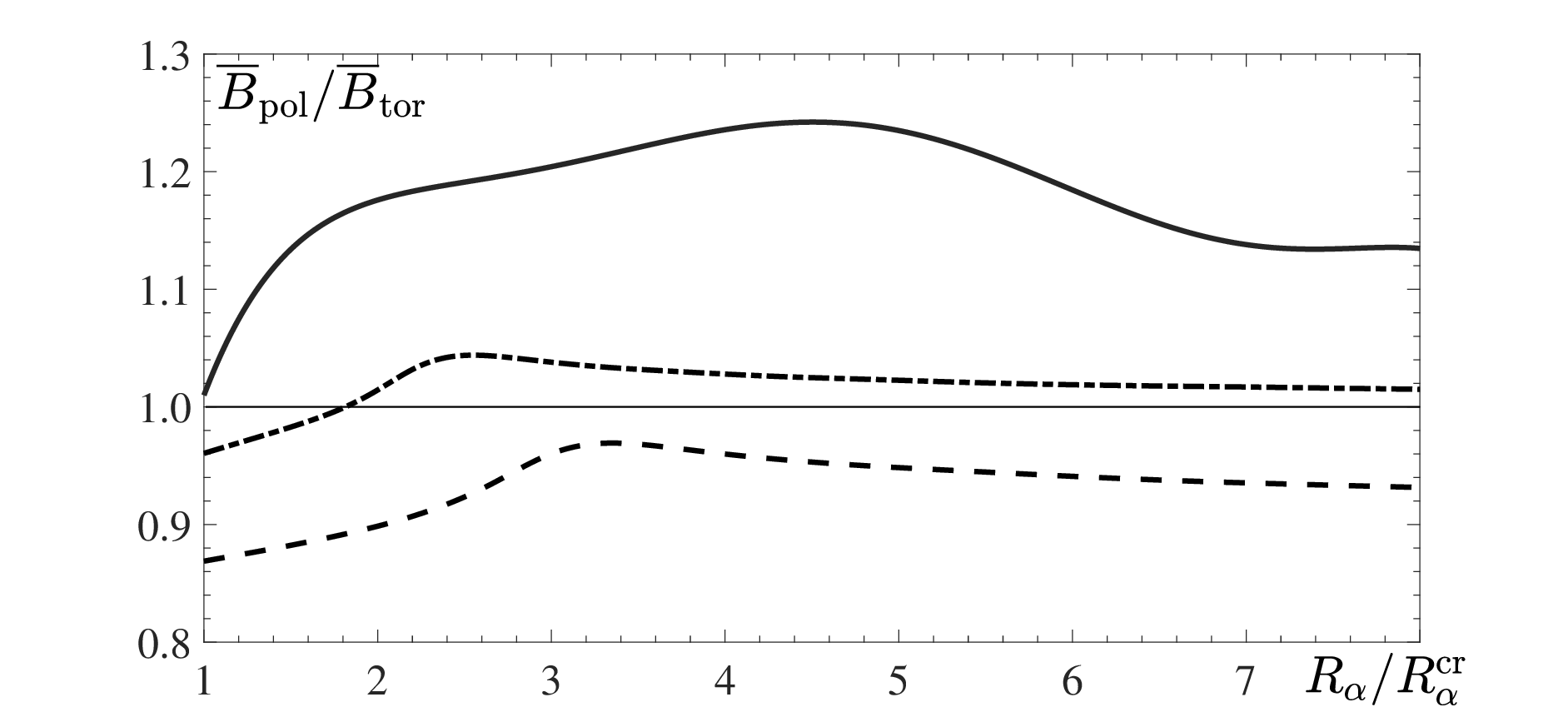}
\caption{\label{Fig14}
The ratio of the maximum values of the poloidal to toroidal mean magnetic fields $\meanB_{\rm pol}/\meanB_{\rm tor}$ versus $R_{\alpha}/R_{\alpha}^{\rm cr}$ for the main sequence stars of three spectral classes: {\rm M}5 (solid); {\rm M}0 (dashed-dotted); {\rm K}5 (dashed),  obtained from numerical simulations of the non-linear $\alpha^2$ mean-field dynamo.
}
\end{figure}

In Fig.~\ref{Fig14} we show the ratio of the maximum poloidal to maximum toroidal mean magnetic field $\meanB_{\rm pol}/\meanB_{\rm tor}$ versus $R_{\alpha}/R_{\alpha}^{\rm cr}$ for the main sequence stars of three spectral classes: {\rm M}5; {\rm M}0 and {\rm K}5,  obtained from numerical simulations in non-linear saturated stage.
It seen that in non-linear saturated stage  $\meanB_{\rm pol}/\meanB_{\rm tor} \sim 1$ .
This is in agreement with equation~(\ref{TM12}) for $R_\omega \ll R_{\alpha} R_{\alpha}^{\rm cr}$
derived for the mean-field $\alpha^2 \Omega$ dynamo.
The latter implies that the generated large-scale magnetic field is caused by the mean-field $\alpha^2 \Omega$ dynamo, whereby the $\alpha^2$ dynamo is modified by a weak differential rotation.

\section{Non-linear theory of axisymmetric $\alpha^2$ dynamo}
\label{sect4}

In this section we discuss a non-linear theory of axisymmetric $\alpha^2$ dynamo.
We consider the axisymmetric mean-field $\alpha^2$ dynamo in spherical coordinates.
The non-linear mean-field induction equation reads
\begin{eqnarray}
{\partial  \over \partial t}
\begin{pmatrix}
A \\ B
\end{pmatrix}
= \left({\hat L} + {\hat N}\right)
\begin{pmatrix}
A \\ B
\end{pmatrix} ,
\label{B2}
\end{eqnarray}
where
\begin{eqnarray}
{\hat L} &=&
\begin{pmatrix}
\Delta_{\rm s} & \alpha_{\rm k} (r ,\theta) \\
- R_\alpha^2 \Delta_{\alpha_{\rm k}}  & \Delta_{\rm s}
\end{pmatrix} ,
\label{B3}
\end{eqnarray}

\begin{eqnarray}
{\hat N} &=&
\begin{pmatrix}
0  &  \alpha_{\rm m}(r ,\theta) \\
- R_\alpha^2 \Delta_{\alpha_{\rm m}} & 0
\end{pmatrix} ,
\label{B4}
\end{eqnarray}
and
\begin{eqnarray*}
\Delta_{\rm s} \tilde \Phi &=& {1 \over r} {\partial^2 \over \partial r^2} (r \tilde \Phi) + {1 \over r^2} {\partial
\over \partial \theta} \biggl({1 \over \sin \theta}
{\partial  \over \partial \theta} (\sin \theta \, \tilde \Phi) \biggr)
\nonumber\\
&\equiv& \left(\Delta - {1 \over r^2 \sin^2 \theta} \right) \tilde \Phi ,
\end{eqnarray*}
and
\begin{eqnarray*}
\Delta_{\alpha_{\rm k,m}} \tilde \Phi &=& {1 \over r} {\partial \over \partial r}
\biggl(\alpha_{\rm k,m} \, {\partial \over \partial r} (r \tilde \Phi) \biggr)
\nonumber\\
&&+ {1 \over r^2} {\partial
\over \partial \theta} \biggl({\alpha_{\rm k,m} \over \sin \theta}
{\partial  \over \partial \theta} (\sin \theta \, \tilde \Phi) \biggr) .
\end{eqnarray*}
Equations~(\ref{B2})--(\ref{B4}) are written in dimensionless variables
(see Section 3).
The operator $\hat L$ describes the kinematic
dynamo. General properties of this operator are outlined in
Appendix~\ref{appendix-B}. Here $\alpha_{\rm k} (r, \theta) = - \alpha_{\rm k}(r, \pi - \theta)$.

We neglect algebraic quenching of the kinetic and magnetic $\alpha$ effect, but take into account the dynamical non-linearity related to the conservation law of the total magnetic helicity for very large magnetic Reynolds numbers.
The magnetic $\alpha$ effect is determined by the evolutionary equation
\begin{eqnarray}
&& {\partial \alpha_{\rm m} \over \partial t} +  {\alpha_{\rm m} \over T_\alpha} = {\bm \nabla} \cdot \left(\kappa_{_{T}} {\bm \nabla} \alpha_{\rm m}\right) - {2 \over \rho} \biggl[(\alpha_{\rm k} + \alpha_{\rm m}) \, R_\alpha^2\, \meanBB_p^{\, 2}
\nonumber\\
&& \quad \quad  \quad - {\hat M} (B,A) + B \, {\partial A \over \partial t} \biggr],
\label{B5}
\end{eqnarray}
where $\meanBB_p = {\rm rot} [A(t,r,\theta) {\bm e}_{\varphi}]$ is the poloidal component of the mean magnetic field, and
\begin{eqnarray*}
&& {\hat M}(B,A) = {\rm rot} \left(B \,{\bm e}_\varphi\right) \cdot {\rm rot} \left(A \,{\bm e}_\varphi\right) .
\end{eqnarray*}
We seek a solution of the non-linear equations~(\ref{B2}) and~(\ref{B5})
in the following form:
\begin{eqnarray}
\begin{pmatrix}
A \\ B
\end{pmatrix}
 = \sum_{n=1}^\infty F^{n}(t) {\bf e}_n (r , \theta), \quad
{\bf e}_n =
\begin{pmatrix}
a_n \\ b_n
\end{pmatrix}
,
\label{B6}
\end{eqnarray}
where ${\bm e}_n$ are the eigenvectors of the operator $\hat L$
for $R_\alpha = R_{\alpha}^{\rm cr}$, i.e., $\hat L^{\rm cr} {\bm e}_n = p_n^{\rm cr} {\bm e}_n$.  Substituting equation~(\ref{B6}) into equation~(\ref{B2})
and taking into account the properties of the eigenvectors given in
Appendix~\ref{appendix-B}, we obtain the following system of equations for the
coefficients $F^\ell(t)$ in equation~(\ref{B6}):
\begin{eqnarray}
{dF^\ell \over dt} - F^\ell p_\ell^{\rm cr} &=& {1  \over 2} \left({dp_\ell \over d\ln R_\alpha}\right)_{\rm cr} \, \, \sum_{n= - \infty}^\infty F^{n}(t) \Big[\alpha^\ell_n
\nonumber\\
&& + R_\alpha^2 \, \tilde \alpha^\ell_n + \left[R_\alpha^2 - (R_{\alpha}^{\rm cr})^2\right] \, G^\ell_n\Big],
\label{B7}
\end{eqnarray}
where $p_\ell^{\rm cr}=p_\ell\left(R_\alpha= R_{\alpha}^{\rm cr}\right)$, and
functions $\alpha^\ell_n$, $\tilde \alpha^\ell_n$ and $G^\ell_n$
are defined by equations~(\ref{B9})--(\ref{B8}) in Appendix~\ref{appendix-B}.

The coefficients $F^\ell(t)$ depend on the non-linearity characterised by $\alpha^\ell_n$.
The equation for $\alpha^\ell_n$ is derived from equation~(\ref{B5}):
\begin{eqnarray}
&& {d\alpha^\ell_n \over dt} + {\alpha^\ell_n \over T_\alpha}   =
- 2\sum_{k,s= - \infty}^\infty F^{k}(t) \biggl\{{dF^s \over dt} S^\ell_{ksn} - F^{s} \biggl[M^\ell_{ksn}
\nonumber\\
&& - R_\alpha^2 \, \left({\alpha^\ell_n \over t_\chi} + K^\ell_{n}\right) \, (\overline{\bm b}_p)_k (\overline{\bm b}_p)_s \biggr]
\biggr\} -  2\kappa_{_{T}} \left(C_1 \alpha^\ell_n - \tilde \alpha^\ell_n \right),
\nonumber\\
\label{B12}
\end{eqnarray}
where $(\overline{\bm b}_p)_n={\rm rot} \left(a_n \,{\bm e}_\varphi\right)$, and
the tensors $M^\ell_{ksn}$, $S^\ell_{ksn}$, and $K^\ell_{n}$ are determined by equations~(\ref{B14})--(\ref{B17}) in
Appendix~A.
The equation for $\tilde\alpha^\ell_n$ is derived from equation~(\ref{B5}) as well:
\begin{eqnarray}
&& {d\tilde \alpha^\ell_n \over dt} + {\tilde \alpha^\ell_n \over T_\alpha} =
- 2\sum_{k,s= - \infty}^\infty F^{k}(t) \biggl\{{dF^s \over dt} \tilde S^\ell_{ksn} - F^{s} \biggl[\tilde M^\ell_{ksn}
\nonumber\\
&& - R_\alpha^2 \, \left({\tilde \alpha^\ell_n \over \tilde t_\chi} + \tilde K^\ell_{n}\right) \, (\overline{\bm b}_p)_k (\overline{\bm b}_p)_s \biggr]
\biggr\} - \kappa_{_{T}} C_2 \tilde \alpha^\ell_n ,
\nonumber\\
\label{BB12}
\end{eqnarray}
where the functions $\tilde M^\ell_{ksn}$, $\tilde S^\ell_{ksn}$, $\tilde K^\ell_{n}$
are determined by equations~(\ref{BB14})--(\ref{BB16}) in
Appendix~\ref{appendix-B}. It is assumed here that the relaxation time $T_\alpha$ of
the magnetic helicity is independent of ${\bf r}$.
Thus the problem reduces to the study of this infinite system of
equations with coefficients determined by the eigenfunctions and
eigenvalues of the linear problem for $R_\alpha = R_{\alpha}^{\rm cr}$.
When $R_\alpha$ in the stellar convective zone  is not much
larger than the critical value $R_{\alpha}^{\rm cr}$ required for the excitation of the dynamo instability,
only few modes are excited.

Let us consider the simplest case, when only one mode is
excited. This is sufficient to estimate the magnitude of the mean
magnetic field in a steady state.
The multi-mode regime could be
considered similarly.
The equations of the single-mode approximation follow from equations~(\ref{B7})--(\ref{BB12}):
\begin{eqnarray}
&& {dF \over dt} - F p^{\rm cr} =
F(t) \Big[\left[R_\alpha^2 - (R_{\alpha}^{\rm cr})^2\right]\, G
+ \alpha + R_\alpha^2 \, \tilde \alpha \Big],
\label{B21}\\
&& {d\alpha \over dt} + {\alpha \over T_\alpha} + \kappa_{_{T}} \left(C_1 \alpha - \tilde \alpha \right) =
- {1 \over 2} \, {dF^2 \over dt} S
\nonumber\\
&& \quad + F^2 \biggl[M - R_\alpha^2 \, \biggl(K+ {\alpha \over t_\chi} \, \overline{\bm b}_p^2 \biggr)\biggr] ,
\label{B22}\\
&& {d\tilde \alpha \over dt} + {\tilde \alpha \over T_\alpha} + \kappa_{_{T}} C_2 \tilde \alpha=
F^2\biggl[\tilde M - R_\alpha^2 \, \biggl(\tilde K
+ {\tilde \alpha \over \tilde t_\chi} \, \overline{\bm b}_p^2 \biggr)\biggr]
\nonumber\\
&& \quad - {1 \over 2} \,{d F^2 \over dt} \tilde S ,
\label{B23}
\end{eqnarray}
The steady-state solution of equations~(\ref{B21})--(\ref{B23}) for this
single-mode approximation yields the
magnitude of the mean toroidal magnetic field near
the stellar surface as
\begin{eqnarray}
 \overline{B} = \left(2 \pi \meanrho_\ast \right)^{1/2} \, {3 \eta_{_{T}} \, \kappa_{_{T}} \over R_\ast} \,
\left({2 G \, T_\alpha \over K} \right)^{1/2} \, f\left[{R_\alpha^2 \over  (R_{\alpha}^{\rm cr})^2 }\right] ,
\label{B24}
\end{eqnarray}
where
\begin{eqnarray}
&& f(X) =  \left({X - 1 \over X-C} \right)^{1/2} \, \left[(R_{\alpha}^{\rm cr})^2 X +2\right]^{-1}
\nonumber\\
&&\quad \times \biggl[1 - 2 G \,  {(R_{\alpha}^{\rm cr})^2 X (X - 1) \over (X-C) \, \left[(R_{\alpha}^{\rm cr})^2 X +2\right]} \, {\overline{\bm b}_p^2 \over t_\chi}
\nonumber\\
&&\quad + \biggl(1 - 4 G \, {(R_{\alpha}^{\rm cr})^2 X (X - 1) \over (X-C) \, \left[(R_{\alpha}^{\rm cr})^2 X +2\right]}
\, {\overline{\bm b}_p^2 \over t_\chi} \biggr)^{1/2}  \biggr]^{-1/2} ,
\label{B25}
\end{eqnarray}
and we consider the case when $K \approx \tilde K$, $M \approx \tilde M$, $K/M=C$,
$t_\chi \approx \tilde t_\chi$ and $C_1=C_2=1$.
The characteristic times $t_\chi$ and $\tilde t_\chi$ are defined by equations~(\ref{B17}) and~(\ref{BB17}) in
Appendix~\ref{appendix-B}.

\begin{figure}
\centering
\includegraphics[width=8.5cm]{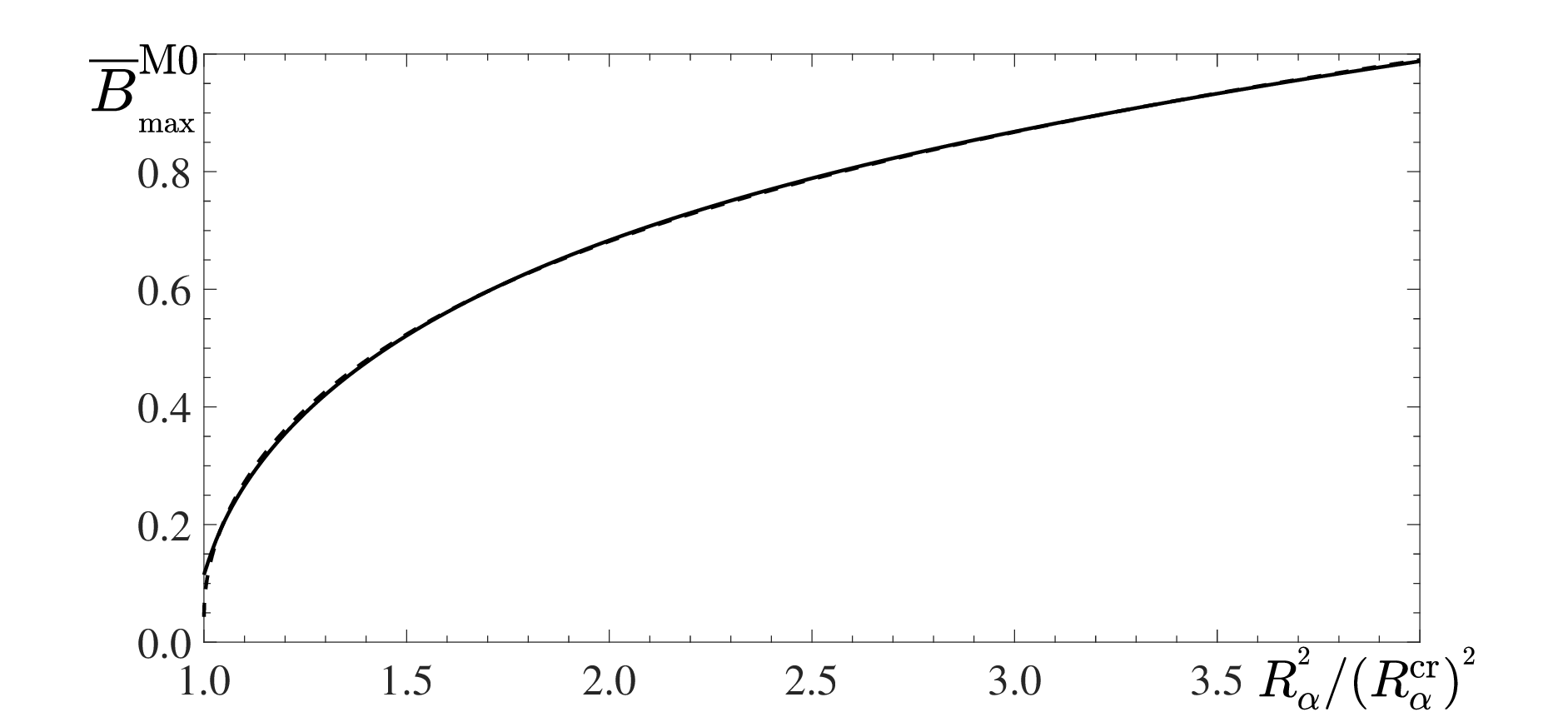}
\caption{\label{Fig15}
The maximum mean magnetic field $\meanB_{\rm max}^{\, \rm M0}$ versus $(R_{\alpha}^2/R_{\alpha}^{\rm cr})^2$ for the main sequence stars of the spectral class {\rm M}0, obtained from numerical simulations  (solid) of the non-linear $\alpha^2$ mean-field dynamo and analytical result (dashed) described by equations~(\ref{B24}) and~(\ref{B25}).
}
\end{figure}

In Fig.~\ref{Fig15} we show the maximum mean magnetic field $\meanB_{\rm max}^{\rm M0}$ (solid) versus $(R_{\alpha}^2/R_{\alpha}^{\rm cr})^2$ for the main sequence stars of the spectral class {\rm M}0 obtained from the mean-field numerical simulations, that is in an agreement with analytical result (dashed) described by equations~(\ref{B24}) and~(\ref{B25}).

\section{Discussions and conclusions}
\label{sect5}

We start the discussion with some comments  related to various numerical simulations.
There are two different kinds of numerical simulations discussed in this paper
which are very important for investigations of the stellar magnetic activity.
The first kind of simulations is direct numerical simulations (DNS) and large eddy simulations (LES)
\citep[see, e.g.,][]{DSB06,BROW08,YCW16,BOV20,KA21,BT22,KBB23}.
The DNS and LES solve exact (or nearly exact) equations
and demonstrate physical effects which mimic the processes occurring inside stars.
For instance, the key feature of the DNS and LES which investigate the stellar magnetic activity, is that they study small-scale
turbulent effects and their influence to large-scale magnetic activity.
However, the used range for the key parameters (e.g., fluid and magnetic Reynolds numbers, P\'{e}clet number,
Rayleigh number, degree of density stratification, etc.) which can be achieved in the DNS and LES, is essentially different from
the parameter range which is typical for stars [see, e.g., review by \cite{KBB23}].

The second type of simulations is mean-field numerical simulations \citep[see, e.g.,][]{CK06,KMS14,SSK15,PI17,PY18}, which are based on various mean-field theories and take into account various turbulence effects by means of turbulent transport coefficients
(e.g., turbulent viscosity, turbulent magnetic diffusivity, turbulent heat conductivity, the alpha effect,
the lambda effect, etc).
The mean-field numerical simulations (MFS) allow to study the large-scale and long-term effects
in spatial and time scales which are much larger
than the turbulent scales.
Mean-field models have been tested using various solar and geophysical observations
of magnetic activity.
Note also that mean-field models are usually improved using the results of the DNS and LES.

In the present paper in addition to the analytical study, we perform non-linear mean-field numerical simulations.
Thus, we can compare our model with others MFS.
In particular, some studies \citep{CK06,KMS14,SSK15}
mainly investigate the kinematic (linear) stage of the mean-field dynamo instabilities, while
in our paper we study non-linear mean-field dynamo instabilities, taking into account
algebraic and dynamic non-linearities.

\cite{PI17} has performed the MFS of the non-linear axisymmetric
and non-axisymmetric $\alpha^2\Omega$ dynamos
of the fully convective star.
However,  the dynamical quenching of the $\alpha$ effect
is determined in this dynamo model by equation for the total magnetic helicity density rather than
that for the evolution of magnetic helicity density of the small-scale field.
The latter is a weak point of this non-linear dynamo model, because the magnetic
$\alpha$ effect is determined by the evolution of the current helicity of the small-scale field,
which is caused by the production and transport of the magnetic helicity density of the small-scale magnetic field.

In the present study, we investigate the non-linear axisymmetric $\alpha^2\Omega$ and $\alpha^2$ dynamos using the dynamic
equation for the evolution of magnetic helicity density of the small-scale field.
We compare the results of the performed MFS with the developed non-linear theory of mean-field dynamo.
In particular, the derived scaling for the magnitude of the mean toroidal magnetic field near
the stellar surface as a function of various key parameters is in agreement with
the results of the performed MFS (see, e.g., Fig.~\ref{Fig15}).

A majority of the observed stars are fast rotating, because
much more easy to observe fast rotating stars generated strong magnetic fields.
Our theoretical study and mean-field numerical simulations suggest that
for fast rotating low-mass main sequences stars with the spectral classes from M5 to G5,
the generated large-scale magnetic field is caused
by the mean-field $\alpha^2\Omega$ dynamo,
where the $\alpha^2$ dynamo is modified by a weak differential rotation.
The latter implies that $R_\omega \ll R_{\alpha} R_{\alpha}^{\rm cr}$.
However, even a weak differential rotation in the non-linear phase of magnetic field evolution
causes drastic changes in magnetic activity,
resulting in chaotic behaviour where long-term evolution (with the period about thousand
years) is accompanied by fast changes of the several tens years.

In view of observations, this multi-timescale system causes very complicated patters in magnetic activity, e.g.,
the fast rotating stars with the same rotation rates and the same spectral classes may have different magnetic activity.
This implies necessity of long-term observational programs of the stellar magnetic activity.
The long-term behaviour of the magnetic activity is related to the characteristic time of
the evolution of the magnetic helicity density of the small-scale magnetic field.
The performed mean-field numerical simulations have shown that
the kinematic and non-linear phases of magnetic field evolution
are very different.
For instance, non-linear effects cause a threshold in the differential rotation that is
necessary for a transition between aperiodic and quasi-periodic regime.
We demonstrate that period of non-linear oscillations decreases with
increase of the differential rotation.

\section*{Acknowledgments}

The work of NK and NS was partially supported by the Russian Science Foundation (grant 21-72-20067).
IR acknowledges the hospitality of NORDITA.

\section*{Data Availability}

\noindent
The data that support the findings of this study are available
from the corresponding author upon reasonable request.

\appendix

\section{The functions $\Psi_{1}(\omega)$ and $\Psi_{2}(\omega)$}
\label{appendix-A}

The functions $\Psi_{1}(\omega)$ and $\Psi_{2}(\omega)$
are given by
\begin{eqnarray*}
\Psi_{1}(\omega) = (3 - \sigma) \, \Big[4 A_1^{(3)}(2\omega) - {3 \over \pi} \bar
A_1(\omega^{2})\Big] -{\lambda\over 2} \Big[4 A_1^{(2)}(2 \omega)
\nonumber \\
- A_1^{(2)}(\omega)\Big] +
3 (\sigma - 1) \Big[4 C_1^{(3)}(2\omega) - {3 \over \pi} \bar
C_1(\omega^{2})\Big] ,
\end{eqnarray*}
\begin{eqnarray*}
\Psi_{2}(\omega) &=& 3 (\sigma - 1)\left[4 C_3^{(3)}(2\omega) - {3 \over \pi} \bar
C_3(\omega^{2})\right].
\end{eqnarray*}
where
\begin{eqnarray*}
A_{1}^{(2)}(\omega) &=& 6 \biggl[{\arctan (\omega) \over \omega}
\biggl(1 + {1 \over \omega^{2}} \biggr) - {3 \over \omega^{2}} +
{2 \over \omega^{3}} S(\omega) \biggr],
\nonumber\\
A_{1}^{(3)}(\omega) &=& J_{0}^{(3)}(\omega) - J_{2}^{(3)}(\omega) ,
\nonumber\\
C_{1}^{(3)}(\omega) &=& {1 \over 4} \biggl[J_{0}^{(3)}(\omega) - 2 J_{2}^{(3)}(\omega) + J_{4}^{(3)}(\omega)\biggr] ,
\nonumber\\
C_{3}^{(3)}(\omega) &=& {1 \over 4} \biggl[- J_{0}^{(3)}(\omega) + 6 J_{2}^{(3)}(\omega) - 5 J_{4}^{(3)}(\omega)\biggr] ,
\end{eqnarray*}
and
\begin{eqnarray*}
\bar A_{1}(\omega^2) = {2 \pi \over \omega^2} \biggl[(\omega^2 + 1) {\arctan
(\omega) \over \omega} - 1 \biggr] ,
\end{eqnarray*}
\begin{eqnarray*}
\bar C_{1}(\omega^2) = {\pi \over 2\omega^4} \biggl[(\omega^2 + 1)^{2} {\arctan
(\omega) \over \omega} - {5 \omega^2 \over 3} - 1 \biggr] ,
\end{eqnarray*}
\begin{eqnarray*}
\bar C_{3}(\omega^2) = - {\pi \over 2\omega^4} \left[\left[\omega^4 + 6 \omega^2 + 5\right]
{\arctan (\omega) \over \omega} - {13 \omega^2 \over 3} - 5 \right] .
\end{eqnarray*}
where
\begin{eqnarray*}
J_{0}^{(3)}(\omega) &=& 2 \biggl[2 {\arctan (\omega) \over \omega}
+ {1 \over \omega^{4}} \ln\left(1 + \omega^{2}\right) - {1 \over \omega^{2}}\biggr],
\end{eqnarray*}
\begin{eqnarray*}
J_{2}^{(3)}(\omega) &=& {6 \over \omega^{2}} \biggl[1 - 2 {\arctan (\omega) \over \omega}
+ {1 \over \omega^{2}} \ln\left(1 + \omega^{2}\right) \biggr],
\end{eqnarray*}
\begin{eqnarray*}
J_{4}^{(3)}(\omega) &=& {2 \over \omega^{2}} \biggl[1 - {3 \over \omega^{2}} \biggl(2 {\arctan (\omega) \over \omega}
- 2
\\
&&+ \ln\left(1 + \omega^{2}\right) \biggr) \biggr] ,
\end{eqnarray*}
and $S(\omega) = \int_{0}^{\omega} [\arctan (y) / y] \,d y$.

In the case of $ \omega \ll 1 $ these functions are given by
\begin{eqnarray*}
A_{1}^{(2)}(\omega) = {8 \over 3} \left(1 - {3 \omega^{2} \over 25}
\right) ,
\end{eqnarray*}
\begin{eqnarray*}
A_{1}^{(3)}(\omega) = 2 \left(1 - {2 \omega^{2} \over 15}
\right) ,
\end{eqnarray*}
\begin{eqnarray*}
C_{1}^{(3)}(\omega) = {2 \over 5} \left(1 - {2 \omega^{2} \over 21}
\right) ,
\quad
C_{3}^{(3)}(\omega) = - {8 \over 105} \, \omega^{2} ,
\end{eqnarray*}
\begin{eqnarray*}
J_{0}^{(3)}(\omega) = 3 \left(1 - {2 \omega^{2} \over 9}
  + {\omega^{4} \over 10} \right) ,
\end{eqnarray*}
\begin{eqnarray*}
J_{2}^{(3)}(\omega) = 1 - {2 \omega^{2} \over 5}
  + {3\omega^{4} \over 14} ,
\end{eqnarray*}
\begin{eqnarray*}
J_{4}^{(3)}(\omega) = {3 \over 5} \left(1 - {10 \omega^{2} \over 21}
  + {5\omega^{4} \over 18} \right) ,
\end{eqnarray*}
In the case of $\omega \gg 1$ these functions are given by
\begin{eqnarray*}
A_{1}^{(2)}(\omega) = {3 \pi \over \omega} -  {24 \over \omega^{2}} ,
\quad A_{1}^{(3)}(\omega) = {2 \pi \over \omega} ,
\end{eqnarray*}
\begin{eqnarray*}
C_{1}^{(3)}(\omega) = {\pi \over 2\omega} , \quad C_{3}^{(3)}(\omega) = -{\pi \over 2\omega} ,
\end{eqnarray*}
\begin{eqnarray*}
J_{0}^{(3)}(\omega) = {2 \pi \over \omega} -  {6 \over \omega^{2}} , \quad
 J_{2}^{(3)}(\omega) = {6 \over \omega^{2}} , \quad  J_{4}^{(3)}(\omega) = {2 \over \omega^{2}} .
\end{eqnarray*}

In the case of $ \omega^{2} \ll 1 $ these functions are given by
\begin{eqnarray*}
\bar A_{1}(\omega^{2}) &\sim& {4 \pi \over 3} \, \left(1 - {\omega^{2} \over 5} \right)  \;,
\quad \bar A_{2}(\omega^{2}) \sim - {8 \pi \over 15} \, \omega^{2} \;,
\\
\bar C_{1}(\omega^{2}) &\sim& {4 \pi \over 15} \, \left(1 - {\omega^{2} \over 7} \right), \quad \bar C_{2}(\omega^{2}) \sim {32 \pi \over 315} \, \omega^{4},
\\
\bar C_{3}(\omega^{2}) &\sim& - {8 \pi \over 105} \, \omega^{2} .
\end{eqnarray*}
In the case of $\omega^{2} \gg 1 $ these functions are given by
\begin{eqnarray*}
\bar A_{1}(\omega^{2}) &\sim& {\pi^{2} \over \omega}  , \quad \bar A_{2}(\omega^{2})
\sim - {\pi^{2} \over \omega} ,
\\
\bar C_{1}(\omega^{2}) &\sim& {\pi^{2} \over 4\omega} - {4 \pi \over 3 \omega^{2}} , \quad
\bar C_{2}(\omega^{2}) \sim {3\pi^{2} \over 4\omega} ,
\\
\bar C_{3}(\omega^{2}) &\sim& - {\pi^{2} \over 4\omega}  + {8 \pi \over 3 \omega^{2}} .
\end{eqnarray*}

\section{Functions for axi-symmetric non-linear $\alpha^2$ dynamo}
\label{appendix-B}

The functions $\alpha^\ell_n$, $\tilde \alpha^\ell_n$ and $G^\ell_n$
entering in equation~(\ref{B7}), are given by
\begin{eqnarray}
\alpha^\ell_n &=& Q_\ell^{-1} \, \int \alpha_{\rm m} \, a^{\ell , \ast} \, b_n \,d^3r ,
\label{B9}\\
\tilde \alpha^\ell_n &=& Q_\ell^{-1} \, \int \alpha_{\rm m} \, {\hat M}(b^{\ell , \ast},a_n) \,d^3r ,
\label{B10}\\
G^\ell_n &=&  Q_\ell^{-1} \, \int \alpha_{\rm k} \,
{\hat M}(b^{\ell , \ast},a_n) \,d^3r ,
\label{B8}\\
Q_\ell &=& \int \alpha_{\rm k} \, a^{\ell , \ast} \, b_\ell \,d^3r .
\label{B11}
\end{eqnarray}

The tensors $M^\ell_{ksn}$, $S^\ell_{ksn}$, and $K^\ell_{n}$ entering in equation~(\ref{B12}) are given by
\begin{eqnarray}
M^\ell_{ksn} &=& Q_\ell^{-1} \, \int {{\hat M}(b_k,a_s)
\over \rho(r)}  \, a^{\ell , \ast} \, b_n \,d^3r ,
\label{B14}
\end{eqnarray}
\begin{eqnarray}
S^\ell_{ksn} &=& Q_\ell^{-1} \, \int {1\over \rho(r)}
 \, a^{\ell , \ast} \, b_n b_k a_s \,d^3r ,
\label{B15}
\end{eqnarray}
\begin{eqnarray}
K^\ell_{n} &=& Q_\ell^{-1} \, \int {\alpha_{\rm k} \over \rho(r)}
 \, a^{\ell , \ast} \, b_n \,d^3r ,
\label{B16}
\end{eqnarray}
\begin{eqnarray}
t_\chi^{-1} &=& Q_\ell^{-1} \,  \int
 \, a^{\ell , \ast} \, b_n \,d^3r .
\label{B17}
\end{eqnarray}

The tensors $\tilde M^\ell_{ksn}$, $\tilde S^\ell_{ksn}$, $\tilde K^\ell_{n}$ entering in equation~(\ref{BB12}) are given by
\begin{eqnarray}
\tilde M^\ell_{ksn} &=& Q_\ell^{-1} \, \int {1
\over \rho(r)} \,  {\hat M}(b_k,a_s)\, {\hat M}(b^{\ell , \ast},a_n) \,d^3r ,
\label{BB14}
\end{eqnarray}
\begin{eqnarray}
\tilde S^\ell_{ksn} &=& Q_\ell^{-1} \, \int {1 \over \rho(r)}
 \, {\hat M}(b^{\ell , \ast},a_n) \, b_k \, a_s \,d^3r ,
\label{BB15}
\end{eqnarray}
\begin{eqnarray}
\tilde K^\ell_{n} &=& Q_\ell^{-1} \, \int {\alpha_{\rm k} \over \rho(r)}
 \, {\hat M}(b^{\ell , \ast},a_n) \,d^3r ,
\label{BB16}
\end{eqnarray}
\begin{eqnarray}
\tilde t_\chi^{-1} &=& Q_\ell^{-1} \, \int {1 \over \rho(r)}
 \, {\hat M}(b^{\ell , \ast},a_n) \,d^3r .
\label{BB17}
\end{eqnarray}

\end{document}